\documentclass{aa} 
\usepackage{natbib}
\usepackage{graphicx}
\usepackage{txfonts}
\begin{document}
   \title{The JCMT \element[][12]{CO}(3--2) Survey of the Cygnus X Region: 
I. A Pathfinder}

   \titlerunning{A Cygnus X \element[][12]{CO}(3--2) Survey Pathfinder}
   \authorrunning{Gottschalk et al.}

   \author{M. Gottschalk \inst{1,2}
          \and R. Kothes \inst{1}
          \and H. E. Matthews \inst{1}
          \and T. L. Landecker \inst{1}        
          \and W. R. F. Dent \inst{3}
        }
	  
   \institute{National Research Council of Canada,
              Herzberg Institute of Astrophysics,
              Dominion Radio Astrophysical Observatory,
              P.O. Box 248, Penticton, British Columbia,
              V2A 6J9, Canada
         \and Department of Physics and Astronomy, University of
              British Columbia, 6224 Agricultural Road, Vancouver, British
              Columbia, V6T 1Z1, Canada
         \and ALMA SCO, Alonso de Cordova 3107, Vitacura, Santiago, Chile
              }

   \date{Received ; accepted }
   \offprints{R. Kothes}

  \abstract
	{Cygnus X is one of the most complex areas in the sky, rich in 
	massive stars; Cyg OB2 (2600 stars, 120 O stars) and other OB 
	associations lie within its boundaries. This complicates 
	interpretation, but also creates the opportunity to investigate
	accretion into molecular clouds and many subsequent stages of 
	star formation, all within one small field of view. Understanding 
	large complexes like Cygnus X is the key to understanding the 
	dominant role that massive star complexes play in galaxies across 
	the Universe.}
	{The main goal of this study is to establish feasibility of a 
	high-resolution CO survey of the entire Cygnus X region by observing part of it
	as a Pathfinder, and to evaluate the survey as a tool for 
	investigating the star-formation process. We can investigate the mass
	accretion history of outflows, study interaction between
	star-forming regions and their cold environment, and examine
	triggered star formation around massive stars.}
	{A 2$\degr\times4\degr$ area of the Cygnus X region has been
	mapped in the \element[][12]{CO}(3--2) line at an angular resolution
	of {15\arcsec} and a velocity resolution of $\sim$0.4~km~s$^{-1}$ using HARP-B and ACSIS on
	the James Clerk Maxwell Telescope. The star formation process is 
	heavily connected to the life-cycle of the molecular material in the 
	interstellar medium. The high critical density
	of the \element[][12]{CO}(3--2) transition reveals clouds in
	key stages of molecule formation, and shows processes that
	turn a molecular cloud into a star.}
	{We observed $\sim$15\% of Cygnus X, and demonstrated that a
	full survey would be feasible and rewarding. We detected three
	distinct layers of \element[][12]{CO}(3--2) emission, related
	to the Cygnus Rift (500-800~pc), to W75N (1-1.8~kpc), and to
	DR~21 (1.5-2.5~kpc). Within the Cygnus Rift, \ion{H}{i} self-absorption
	features are tightly correlated with faint diffuse CO emission, while
	HISA features in the DR~21 layer are mostly unrelated to any
	CO emission. 47 molecular outflows were detected in the Pathfinder, 
	27 of them previously unknown. Sequentially triggered star formation 
	is a widespread phenomenon.}
   {}


   \maketitle
%

\section{Introduction}

Star formation is the process that builds galaxies, and the formation
of massive stars is the dominant influence on their evolution: massive
stars trigger the formation of further stars and rapidly enrich their
environment through stellar winds and supernova explosions. In the
Milky Way, W49A, Westerlund 1 and Cygnus X have been identified as
sites where large clusters of massive stars have formed.  W49A
contains at least 100~O~stars in four clusters \citep{alve03} embedded
in some $10^{6}{{\rm{M}}_{\odot}}$ of molecular gas. Westerlund 1 has
at least 200 cluster members \citep{clar05} including a rich
population of Wolf-Rayet stars and Luminous Blue Variables. Cygnus X
is equally a laboratory where we can study the formation of massive
stars: the region Cyg OB2, originally classified as an OB association,
is actually a very large star cluster that contains about 120 O stars
\citep{knod00,knod04} and is within the boundaries of Cygnus X at
a distance of about 1.7~kpc. Other
OB associations at similar distances, Cyg OB1, OB8, and OB9 lie within
the Cygnus X boundary.  Cygnus X is relatively nearby, much
closer than Westerlund 1 (3.9~kpc -- \citet{koth07}) or W49A (11.4~kpc
-- \citet{alve03}). The mass of molecular gas in Cygnus X is estimated to be
${\sim}4.7{\times}10^{6}{{\rm{M}}_{\odot}}$ \citep{schn06} assuming
all the molecular gas is at the distance determined for Cyg~OB2.

The study of Cygnus X as an entity is almost as old as radio astronomy
itself. It was first recorded by \citet{pidd52} as an extended source
in the vicinity of Cygnus A, an object of much interest at the time;
the ``X'' denoted the extended nature of the source. These early 
measurements were made with an angular resolution of several degrees, 
but \citet{pidd52} determined a size for the source that matches modern 
estimates and also correctly deduced the thermal nature of the radio emission.
With gradually improving angular resolution, the extended source broke
up into sub-components. For example, \citet{down66} identified 26
discrete sources within Cygnus X with a $10.8'$ beam. Eventually
arcminute resolution was reached, $2.6'$ at 4.8~GHz \citep{wend84},
$4'$ at 408~MHz \citep{wend91}, and $1'$ in the Canadian Galactic
Plane Survey at 1420~MHz \citep{tayl03}. 

Optical obscuration is high in this direction, because Cygnus X lies partly
behind the Cygnus Rift (at $\sim$700~pc). Radio observations, of
course, reveal the individual objects behind this obscuration, but compress the
``bewildering amount of detail'' \citep{wend84} into a two-dimensional
image. Observations of radio spectral lines associated with various
Cygnus X objects -- e.g. \citet{piep88} -- give a variety of systemic
velocities, the most negative implying distances up to
4~kpc. Unfortunately kinematic distances within 2 or 3~kpc are
unreliable in this direction because the radial velocity gradient is
near zero, considerably smaller than the typical velocity dispersion
of the gas. All that can be said is that there are many local objects
in Cygnus X, as well as others at larger distances corresponding to the
Perseus arm and even the Outer arm. Since this direction corresponds
to the line of sight along the local arm, the Orion spur, the
suggestion naturally arose that the apparent complexity and large
amount of ionized gas are the simple consequence of viewing the Local arm
end-on \citep{wend91}.

The infrared waveband provided another window through the obscuration,
and the improving technology of IR imaging yielded new information.
\citet{knod00} used 2MASS data to demonstrate that the Cyg OB2
association is much larger than previously believed, containing about
2600 OB stars. This discovery suggested a much greater concentration
of Cygnus X around the distance of Cyg OB2 at 1.7~kpc, and not 
stretched out along the Local arm.  \citet{knod04}
provides a comprehensive review of the Cygnus X region based on
multi-wavelength data.

We now cross another technological boundary. In this paper we present
observations of part of the Cygnus X region with an angular resolution
of $15''$ in the \element[][12]{CO}(3--2) line using the HARP/ACSIS
multi-beam system on the JCMT. We surpass the angular resolution of
the earlier comprehensive line survey of \citet{schn06} by a factor of
12.  Our observations are a Pathfinder for a survey in the
\element[][12]{CO}(3--2) line that could effectively and
systematically map molecular clouds and diffuse emission in this
entire region, revealing the dynamical relationship of various
objects, with sufficient angular resolution to be able to discern the
detailed structure of individual objects and complexes. Our effective
physical resolution is 0.1~pc at a distance of 1.7~kpc.

This paper presents the data obtained from the Pathfinder observations
and gives some examples of the types of objects that can be
studied and the types of questions that could be answered using
the $^{12}$CO(3--2) line, but does not include an exhaustive
discussion of any single process. Subsequent papers will incorporate
further observations and will more thoroughly discuss individual
processes and objects.

\section{Specific Objectives of the JCMT Cygnus X $^{12}$CO(3-2) Survey}

Cygnus X contains hundreds of \ion{H}{ii} regions, IRAS point sources,
stellar clusters and OB associations as well as multiple objects with
very bright radio emission. Here we can probe, within a small 
field of view, a major part of the cycle
of star formation, a process that is strongly connected to the life-cycle 
of the molecular material in the interstellar medium (ISM). We can begin a study
of the condensation of atoms into molecules, the transformation of
dense molecular cloud cores into protostellar objects, the impact of
newly formed and evolved stars on the surrounding interstellar medium,
and, possibly, the triggering of another round of star formation.

\citet{gibs05} investigated \ion{H}{i} self absorption (HISA) features using
data from the Canadian Galactic Plane Survey \citep[CGPS,][]{tayl03}. HISA
clouds often have no CO counterparts, and \citet{gibs05} conclude that
these clouds trace atomic gas on the path to formation of molecular
hydrogen where CO molecules have not yet formed. With our high
sensitivity and resolution we can probe HISA features for CO
emission.  We will eventually have a complete sample inside a very
large molecular cloud and star formation complex, and we will be able
to quantify the HISA-CO relation by comparing relative abundances of
each species. Models of molecule formation rates, such as those of
\citet{glov09}, can determine ages, and give us deeper insights into
the molecule formation process.

Among the many objectives of such a survey is the study of molecular
outflows; the gravitational collapse of a molecular cloud ejects
material in bipolar jets, and the discovery of a bipolar outflow,
along with a dusty disk or envelope, signifies the presence of a protostar.
Studies of outflows
provide a fossil record \citep{ball07} of the mass
accretion. \citet{lada85} outlined a method for calculating dynamical
timescales assuming an average velocity over the lifetime and
analyzing the spatial extent of the lobes. This analysis requires high
angular resolution. For CO-HISA clouds containing outflows we can
compare the outflow's dynamical timescale to the molecule formation
timescale, determining star forming efficiency.  $^{12}$CO(3-2) line
observations can provide basic outflow parameters. The high critical density of
$^{12}$CO(3-2) makes it an excellent tracer of dense material and
line observations with the JCMT are probably the most
sensitive probe of the warm outflow gas. Previous outflow
searches have been biased toward near-infrared signposts of star
formation. It is therefore of vital importance to make an unbiased
and complete search for molecular outflows in a large molecular cloud and
star formation complex such as the Cygnus X region.

Cometary nebulae are common within the Cygnus~X region. Most of the
cometary clouds should contain protostars, which are starbearing
evaporated gas globules (EGGs). Most of the O and B stars within
Cygnus~X belong to OB associations such as Cyg OB2 \citep{knod00} or
related clusters \citep{ledu02}. A complete sample of EGGs with
velocity information will allow us to explore the conditions for
triggered star formation and determine whether evaporation around the
protostar is stunting the growth of what would have been massive
stellar objects. Schneider et al. (2006) proposed that nearly all
molecular clouds in Cygnus X form groups that are dynamically
connected, and that most of the Cygnus X objects are located at the
distance of the Cyg OB2 cluster, 1.7~kpc. Unveiling the sequence of
star formation and the entire evolutionary path of the Cygnus-X
population will test the Schneider et al. postulate; to fulfill this
aim we have to observe the entire Cygnus-X region at sub-arcminute
angular resolution.

\section{Pathfinder Observations and Data Reduction}

\begin{table*}{tb}
\caption{\label{tab:obs} Centre positions, dates of observation, and final sensitivity of the
eight Cygnus X fields observed for the Pathfinder.}
\begin{center}
\begin{tabular}{|c|cccc|} \hline
\bf Field & \bf R.A.(J2000) & \bf DEC(J2000) & \bf Dates (YYYYMMDD) & \bf Noise [mK] \\ \hline
1 & 20$^h$\,25$^m$\,43.7$^s$ & 41$^\circ$\,54\arcmin\,44\arcsec & 20070820 & 500 \\
2 & 20$^h$\,25$^m$\,40.9$^s$ & 42$^\circ$\,54\arcmin\,42\arcsec & 20070819 & 850 \\
3 & 20$^h$\,31$^m$\,05.9$^s$ & 41$^\circ$\,54\arcmin\,44\arcsec & 20070819 & 850 \\
4 & 20$^h$\,31$^m$\,08.2$^s$ & 42$^\circ$\,54\arcmin\,42\arcsec & 20070728 & 750 \\
5 & 20$^h$\,36$^m$\,28.1$^s$ & 41$^\circ$\,54\arcmin\,44\arcsec & 20090714~~~20100716~~~20100802 & 300 \\
6 & 20$^h$\,36$^m$\,35.5$^s$ & 42$^\circ$\,54\arcmin\,42\arcsec & 20100717~~~20100718~~~20100802 & 250 \\
7 & 20$^h$\,41$^m$\,50.3$^s$ & 41$^\circ$\,54\arcmin\,44\arcsec & 20100803 & 300 \\
8 & 20$^h$\,42$^m$\,02.8$^s$ & 42$^\circ$\,54\arcmin\,42\arcsec & 20100725 & 300 \\
\hline
\end{tabular}
\end{center}
\end{table*}

The observations for the Cygnus~X Pathfinder were carried out using
the heterodyne array receiver system \citep{buck09} on the James Clerk
Maxwell Telescope (JCMT), Mauna Kea, Hawaii. This system is composed
of the 16-pixel array, HARP-B, as the frontend receiver and the ACSIS
digital autocorrelation spectrometer as the backend correlator
system. The observations were centered on the 345.7959~GHz line of the
\element[][12]{CO}(J=3--2) transition in single-sideband mode with a
total bandwidth of 1000~MHz and a channel spacing of 0.488~MHz. This
resulted in 2048 channels with a channel spacing of
0.423~km~s$^{-1}$ and a total velocity coverage of about
850~km~s$^{-1}$.  The individual beams of the HARP receiver have half
width of 15\arcsec (FWHM), and the beams are spaced {30\arcsec} apart
on a $4 \times 4$ grid.  The data were obtained using the fast-raster
scanning mode resulting in {7.5\arcsec} sampling in the scanning
direction. For this purpose the 4x4 array is rotated so that the 16
receptors project scans onto the sky which are {7.5\arcsec}
apart. During all of the observations one or two array elements were
out of commission or missing, resulting in gaps in the raw data cubes.
To compensate more easily for the missing receptors individual scans
were separated by half an array.

\begin{figure*}[tb]
\centerline{\includegraphics[bb=15 15 569 555,width=10.5cm,clip]{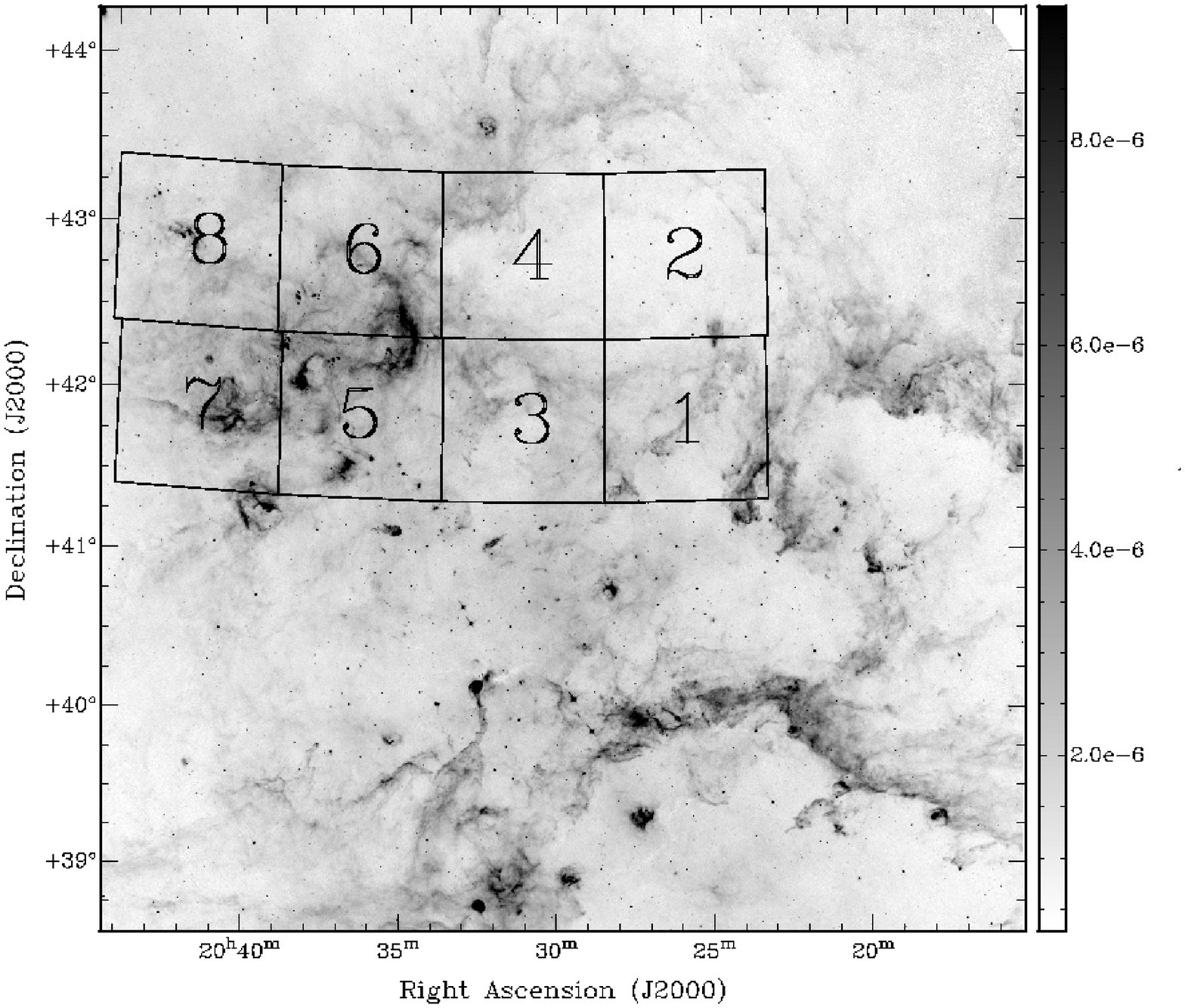}}
\centerline{\includegraphics[bb=15 15 569 555,width=10.5cm,clip]{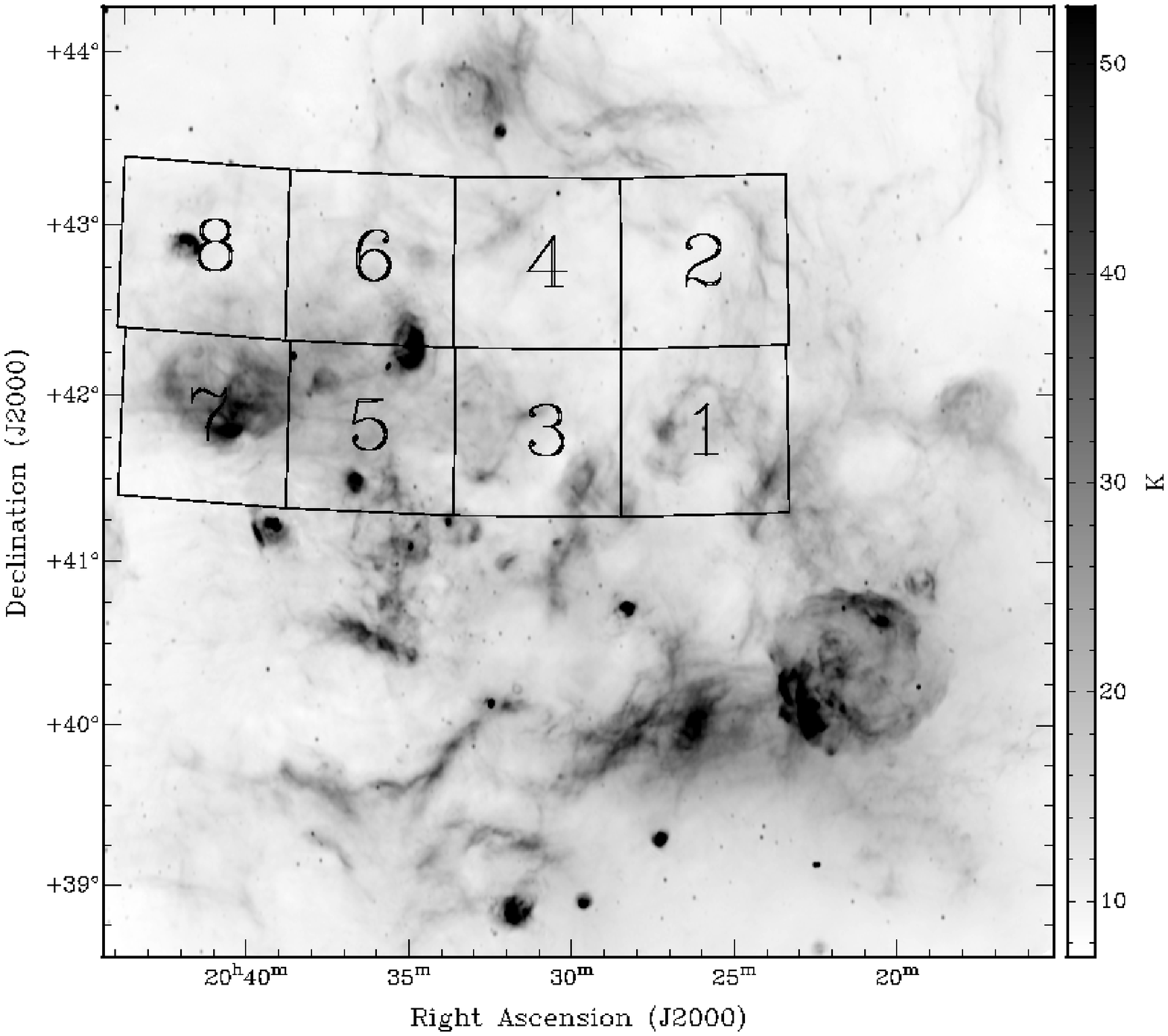}}
\caption{\label{fig:obs} The Cygnus X region as seen in the Midcourse
Space Experiment (MSX) at 8\,$\mu$ (top) and in the Canadian Galactic
Plane Survey \citep{tayl03} in total intensity at 1420~MHz
(bottom). The location of the Pathfinder area with its eight
$1^\circ\times 1^\circ$ fields is indicated.}
\end{figure*}

The final data cube is centered on
$\alpha$(J2000)~=~{20$^{\mathrm{h}}$~34$^{\mathrm{m}}$} and
$\delta$(J2000)~=~{42\degr~26\arcmin} and is a mosaic of eight smaller
fields, each about one square degree in size. The location of those
fields within the Cygnus X region is displayed in Fig.~\ref{fig:obs}
and their characteristics are summarized in Table~\ref{tab:obs}. The
scanning speed of the telescope was {37.5\arcsec} per second with a
5~Hz sampling rate, completing each observation in around 90 minutes.
All fields were scanned at least twice in a boustrophedon pattern,
once along right ascension and once along declination, to reduce
scanning effects. The observations took place in the period May to
August 2007 for fields 1 to 4, July 2009 for field 5 and July to
August 2010 for fields 5 to 8 (see Table~\ref{tab:obs}). The fields
were observed during varying grades of weather, a fact reflected in
significant difference of sensitivity (see Table~\ref{tab:obs}).

Each field was reduced according to the ACSIS DR workflow by using
\textsc{kappa, smurf, convert} and \textsc{gaia} from the STARLINK
package. The full mosaic containing all the fields was assembled using
\textsc{madr} from the DRAO Export Software Package. The data were
originally in the form of time-stamped streams of spectra for each
detector. These time series were cleaned of any noisy spectral bounds,
any bad/noisy receptors, and any spikes amongst the spectra. The
cleaned raw data from each scanning direction were then combined into
one data cube and analyzed simply, only to determine ranges free of
emission lines for baseline correction. A second-order baseline was
first fitted to the emission-line-free channels along the spectral
axis and subtracted from the data cube.

Due to atmospheric condition and receiver instabilities, base levels
of individual scans are not always constant. These ``scanning
effects'' are a common feature of raster scanning. They were
removed by using the unsharp masking technique as described in
\citet{sofu79}. The assumption for this method is that the baseline
variation along each scan is a smooth function and independent of its
neighbours. For the Pathfinder data we used a filter beam of
$90\arcsec \times 30\arcsec$ ($\Theta_{maj} \times \Theta_{min}$,
where $\Theta_{maj}$ is the beam perpendicular to the scanning
direction and $\Theta_{min}$ the beam parallel to it). The convolution
of the original data set with this filter beam effectively smears out
the scanning effects. Therefore a difference map of the smoothed data
subtracted from the original data includes only information about the
baseline error and structures smaller than the filter beam. The latter
are distributed randomly around the smooth function representing the
scanning effect. A polynomial of order 3 is then fitted to the
difference data after subtracting all left-over emission above
$5\sigma$. The fitted baselines are then removed from the original
dataset resulting in a ``true'' distribution of the $^{12}$CO(3-2) emission. 

Occasionally there were holes in the clean data cube where missing
receptors on HARP overlapped from each scanning direction. These were
interpolated using the \textsc{kappa} \emph{fillbad} routine which
averages the surrounding pixels. After combining the cubes for the
different scanning directions a clean data cube for each field was
obtained.  The clean data cube was then ready to be combined with the
other cubes to create the full mosaic.

Emission in the final data cube was contained in a region of 356 channels. To reduce its size, 
only these channels were included in the final data product. The resulting cube is
$\sim2\degr\times 4\degr$, with an angular resolution of $15\arcsec$, and spans a velocity
range of $-100$~to~$+50$~km~s$^{-1}$. 
Channels at velocities less than $-100$~km~s$^{-1}$ and greater than $50$~km~s$^{-1}$ were 
inspected for emission of extragalactic origin, but none was found.

\section{The Pathfinder Data}

\begin{figure*}
\centerline{
\includegraphics[bb=20 65 540 590,width=16cm,clip]{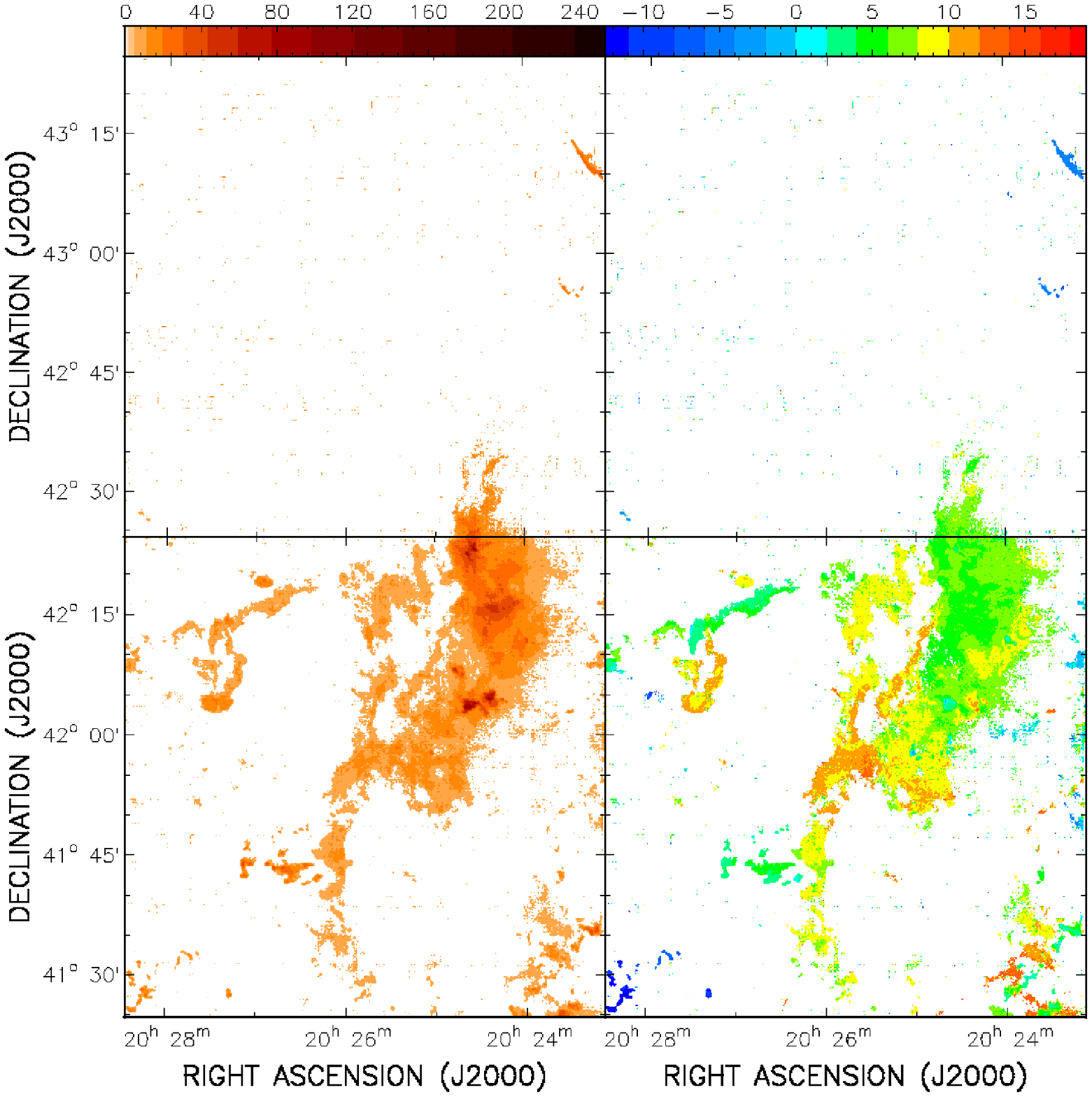}}
\caption{\label{fig:fl12} Zeroth and first moment maps of field 1 (bottom) and field 2 (top).
For the calculation of zeroth and first moment maps, please consult the text in section 4.
Both maps were integrated from $-20$~km\,s$^{-1}$ to $+30$~km\,s$^{-1}$. Only velocities 
with a signal above 3$\sigma$ were used in the integration. In the images only those pixels
are displayed that have a signal above 3$\sigma$ in at least 3 velocity channels. The zeroth 
and first moment maps are displayed in units of K\,km\,s$^{-1}$ and km\,s$^{-1}$, respectively.}
\end{figure*}

\begin{figure*}
\centerline{
\includegraphics[bb=20 65 540 590,width=16cm,clip]{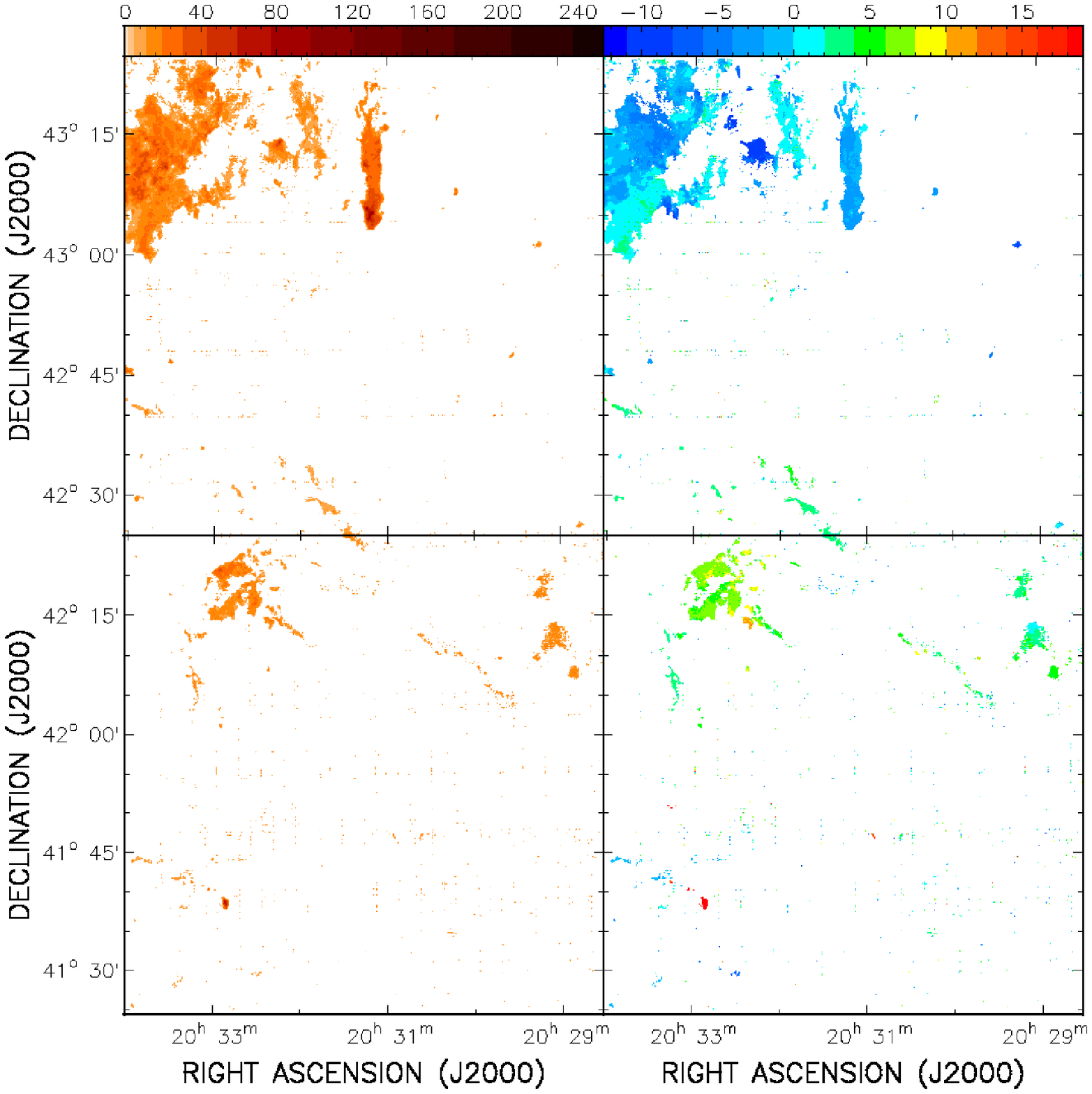}}
\caption{\label{fig:fl34} Zeroth and first moment maps of field 3 (bottom) and field 4 (top).
For the calculation of zeroth and first moment maps, please consult the text in section 4.
Field 3 was integrated from $-12$~km\,s$^{-1}$ to $+25$~km\,s$^{-1}$ and field 4
from $-15$~km\,s$^{-1}$ to $+20$~km\,s$^{-1}$. In the images only those pixels
are displayed that have a signal above 3$\sigma$ in at least 3 velocity channels. The zeroth 
and first moment maps are displayed in units of K\,km\,s$^{-1}$ and km\,s$^{-1}$, respectively.}
\end{figure*}

\begin{figure*}
\centerline{
\includegraphics[bb=20 65 540 590,width=16cm,clip]{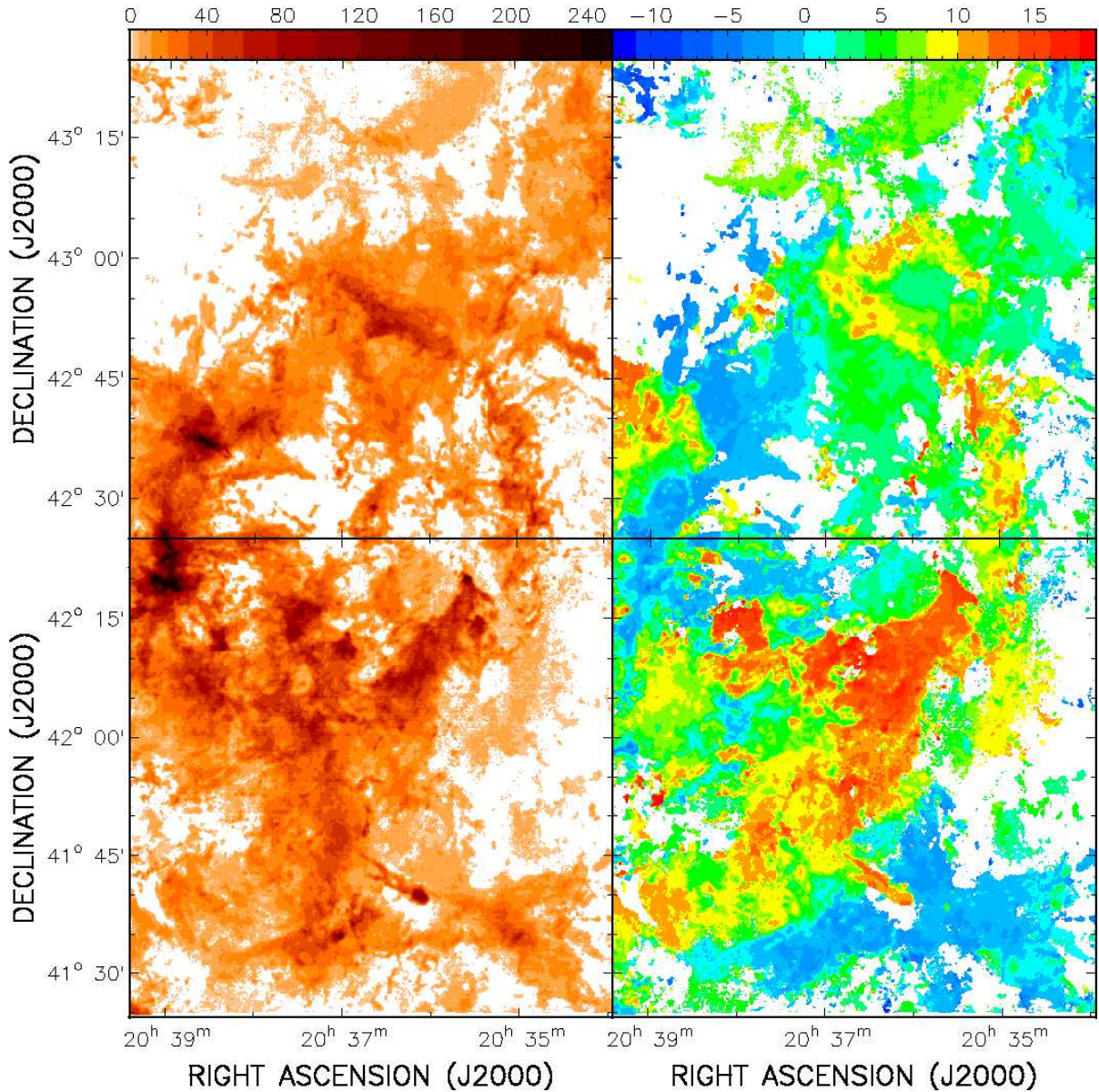}}
\caption{\label{fig:fl56} Zeroth and first moment maps of field 5 (bottom) and field 6 (top).
For the calculation of zeroth and first moment maps, please consult the text in section 4.
Both fields were integrated from $-20$~km\,s$^{-1}$ to $+30$~km\,s$^{-1}$. In the images only those pixels
are displayed that have a signal above 3$\sigma$ in at least 3 velocity channels. The zeroth 
and first moment maps are displayed in units of K\,km\,s$^{-1}$ and km\,s$^{-1}$, respectively.}
\end{figure*}

\begin{figure*}
\centerline{
\includegraphics[bb=20 65 540 590,width=16cm,clip]{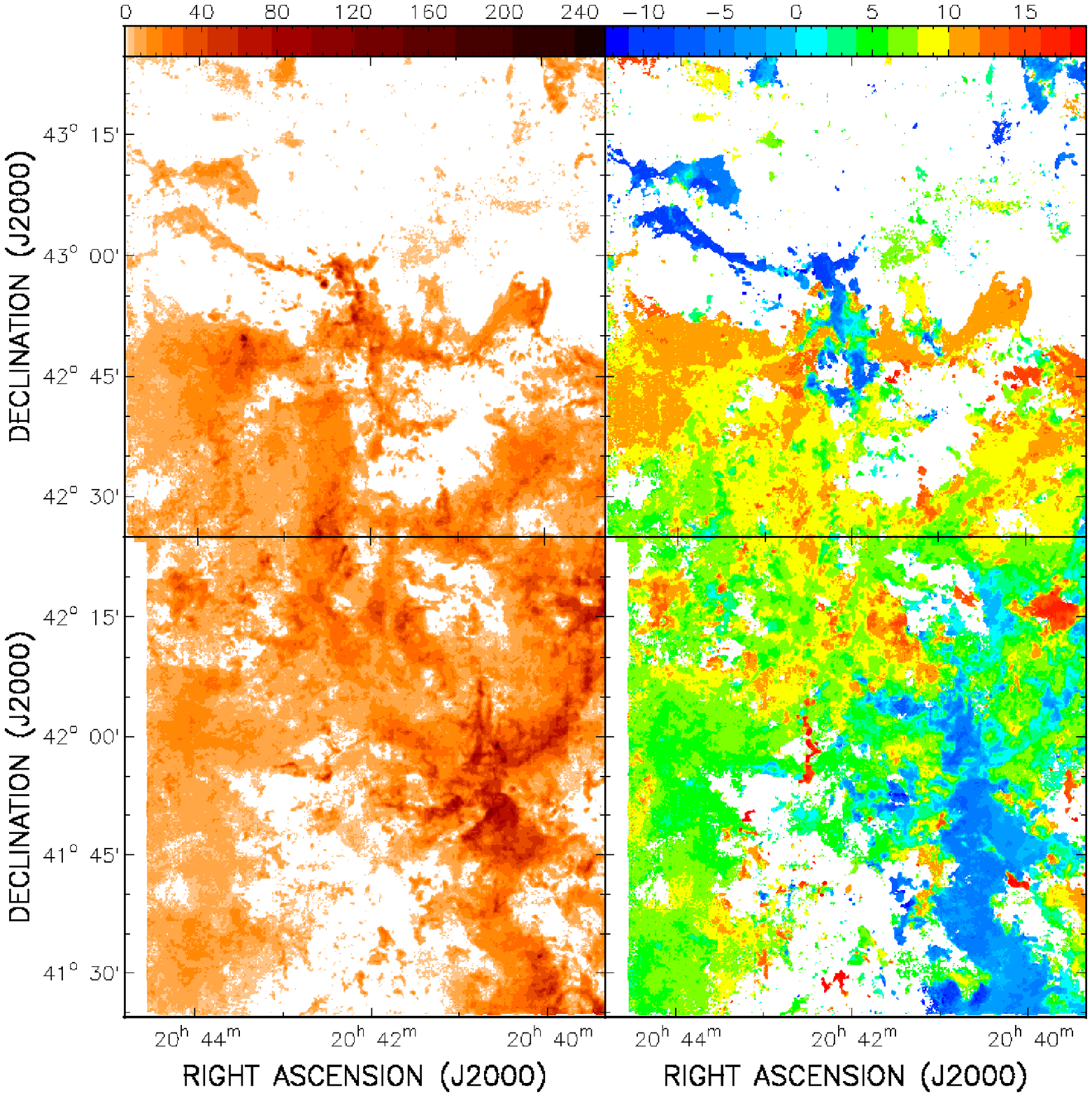}}
\caption{\label{fig:fl78} Zeroth and first moment maps of field 7 (bottom) and field 8 (top).
For the calculation of zeroth and first moment maps, please consult the text in section 4.
Field 7 was integrated from $-18$~km\,s$^{-1}$ to $+25$~km\,s$^{-1}$ and field 8
from $-15$~km\,s$^{-1}$ to $+22$~km\,s$^{-1}$. In the images only those pixels
are displayed that have a signal above 3$\sigma$ in at least 3 velocity channels. The zeroth 
and first moment maps are displayed in units of K\,km\,s$^{-1}$ and km\,s$^{-1}$, respectively.}
\end{figure*}

The observations were taken in two very distinct areas of the Cygnus X region. 
The western part of the Pathfinder covers fields 1 through 4
(see Figs~\ref{fig:obs}, \ref{fig:fl12}, and \ref{fig:fl34}), 
and represents a very ``quiet'' area largely free of dynamical activity. The
eastern part covers fields 
5 through 8 (see Figs~\ref{fig:obs}, \ref{fig:fl56}, and \ref{fig:fl78}) and contains
a lot more activity in the form of \ion{H}{ii} regions and star forming cores, centered at
the well-known complex related to DR\,21. The Pathfinder area was chosen this way to include
both presumably active and less active areas of the Cygnus X region.

We found an astonishing richness in structure and brightness, from
faint, diffuse clouds to dense cores at the resolution limit in both
regions, the quiet western part and the more active eastern part.
Each observed field is shown through a zeroth and first moment map in
Figs.~\ref{fig:fl12} to \ref{fig:fl78}. The zeroth moment maps are
created by integrating the brightness temperature $T_b$ over the
velocity range indicated in the figure captions: $\int T_b\,dv$.  The
first moment maps are created by averaging the velocity $v$ of each
spectrum weighted by its brightness temperature: $\frac{\int
T_b\,v\,dv}{\int T_b\,dv}$. The zeroth moment maps display all the
emission within the velocity range while the first moment maps
illustrate the dominant velocity of the CO emission along each line of
sight.

\subsection{The Western Region}

The western region (Figs~\ref{fig:fl12} and \ref{fig:fl34}) contains
considerably less \element[][12]{CO}(3--2) emission than the eastern
region, as can be seen in a comparison of Figs.~\ref{fig:fl12} through
\ref{fig:fl78}. The bulk of the CO emission in this area seems to
originate from molecular clouds associated with the \ion{H}{ii} region
complex NGC\,6914. Those can be found around 20$^h$25$^m$,
42$^{\circ}$23$'$ in a velocity range from about 0
to +10~km\,s$^{-1}$. Within these clouds we detected four molecular
outflows (see Table~\ref{tab:of}), all of which are accompanied by
IRAS point sources indicating young protostars.  The extended
\element[][12]{CO}(3--2) emission coincides with a HISA complex
discovered in data of the Canadian Galactic Plane Survey by
\citet{gibs05}. This diffuse cloud also completely absorbs the
emission from the bipolar outflow related to IRAS~20227+4154 in the
velocity range $+4.5$ to $+6.4$~km\,s$^{-1}$ indicating a very large
optical depth and so a high molecular column density for that
cloud. This diffuse cloud must be a foreground object to
IRAS~20227+4154, since it absorbs emission from both outflow lobes and
therefore is probably unrelated.

Another 
source of considerable \element[][12]{CO}(3--2) emission in the 
western region is a cometary shaped nebula with CO emission at low negative
velocities (Fig~\ref{fig:fl34}), visible at
$\alpha$(J2000): {20$^{\mathrm{h}}$~31$^{\mathrm{m}}$~11$^{\mathrm{s}}$} and 
$\delta$(J2000) from {43\degr~00\arcmin} to {43\degr~15\arcmin}. Two molecular outflows are found 
towards the dense core within this bright nebulous object. Both outflows were previously unknown.
The position of these outflows
corresponds to the location of IRAS\,20294+4255 indicating that 
protostars are likely present. Due to the low resolution of the IRAS data it is not clear
whether this IRAS source corresponds to one of the outflows or may be a combination of
both.

In the north-east of the western region is another diffuse cloud of
emission at low negative velocities (Fig~\ref{fig:fl34}), which seems
to be unrelated to the cometary feature. These clouds continue into
the eastern region and seem to be part of the very large
molecular cloud complex connected to DR\,21.  The relative lack of
activity within this region is probably due to the absence of
large \ion{H}{ii} regions, typically seen as radio bright DR or
W objects, which have a pronounced effect on their surrounding ISM.

\subsection{The Eastern Region}

\begin{figure*}
\centerline{
\begin{minipage}{7.5cm}
\includegraphics[bb=85 26 695 568,height=6.1cm,clip]{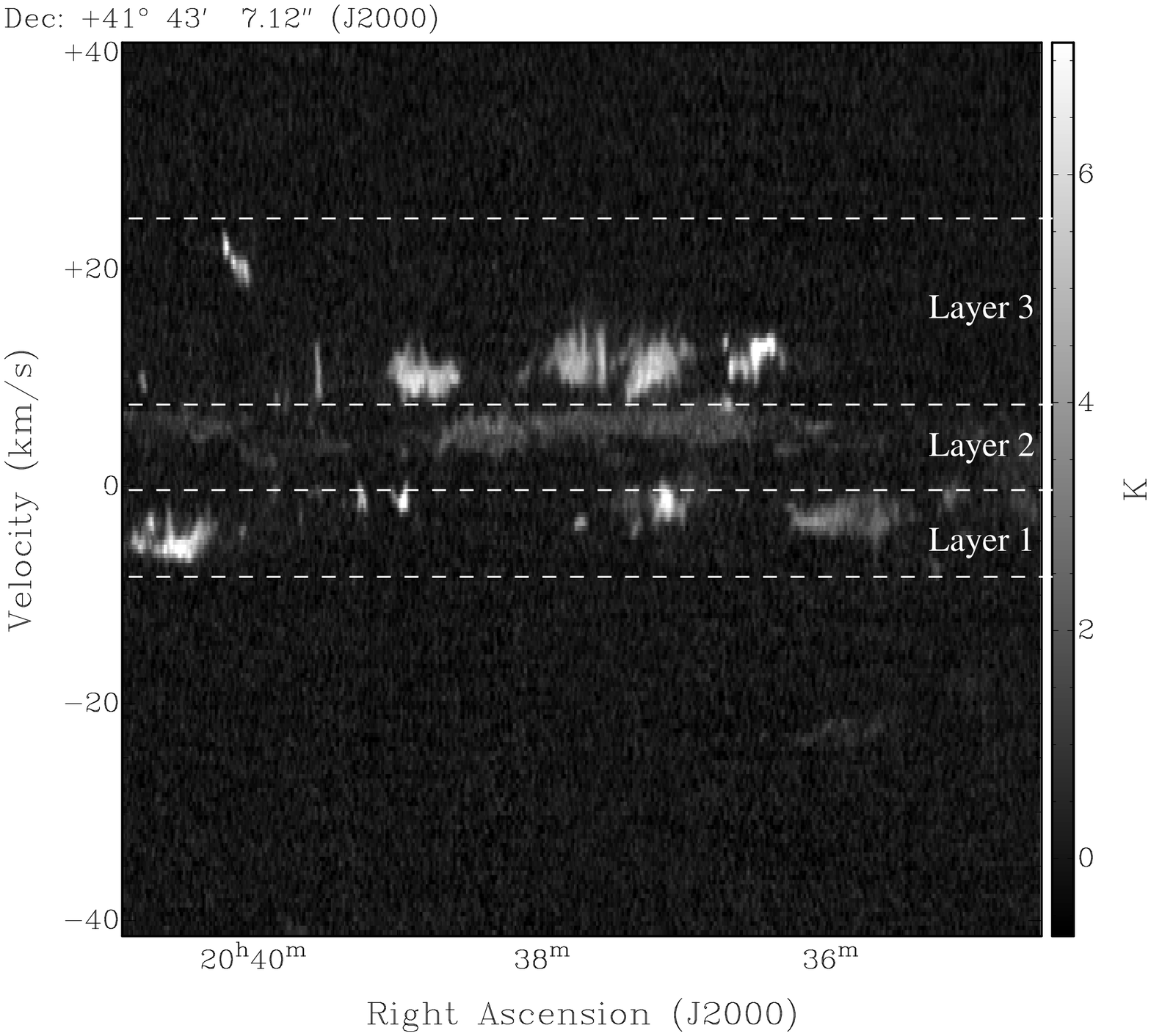}
\end{minipage}
\begin{minipage}{7.5cm}
\includegraphics[bb=75 17 680 552,height=6.1cm,clip]{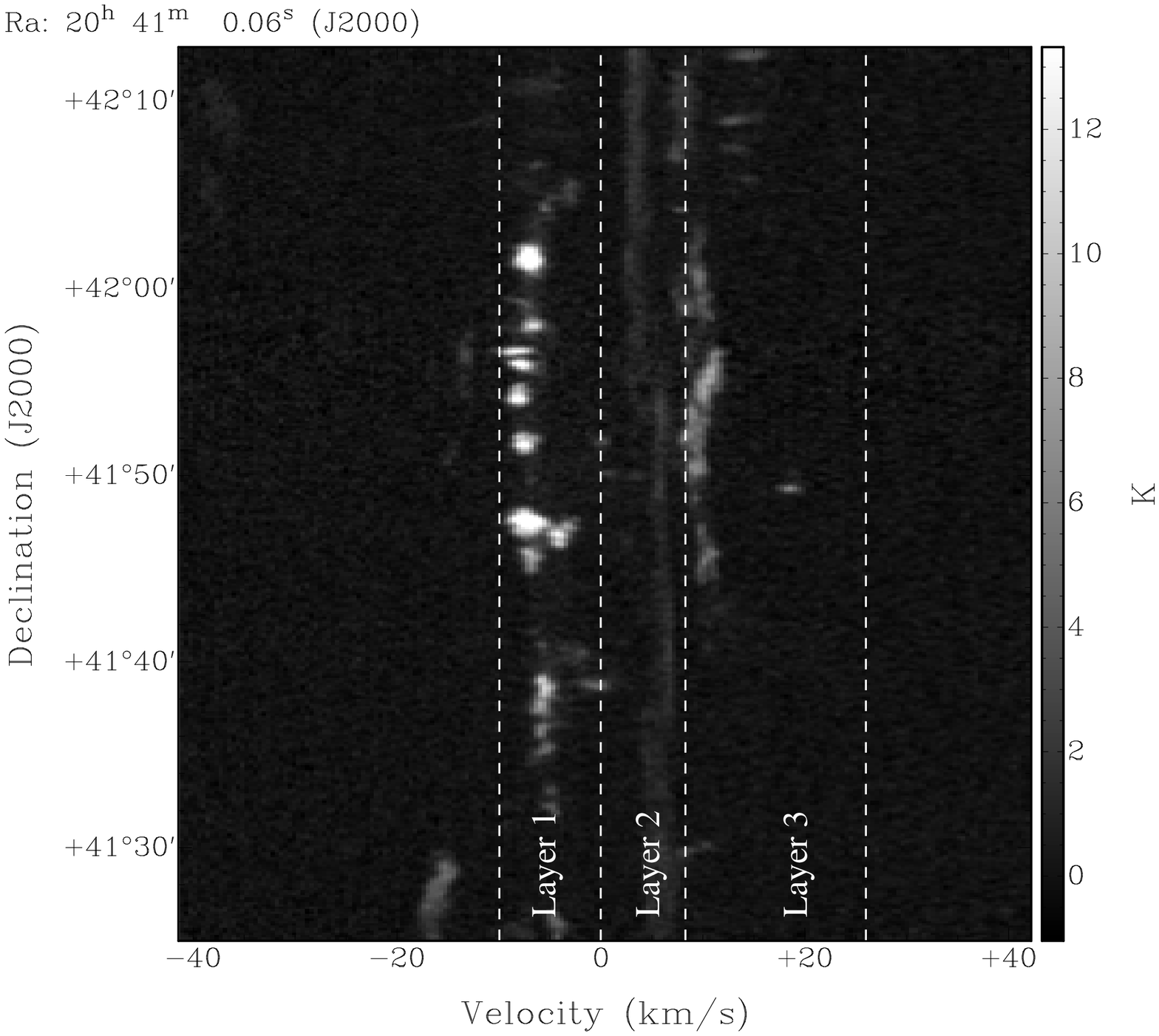}
\end{minipage}
}
\centerline{
\begin{minipage}{7.5cm}
\includegraphics[bb=25 118 570 725,width=6.4cm,angle=-90,clip]{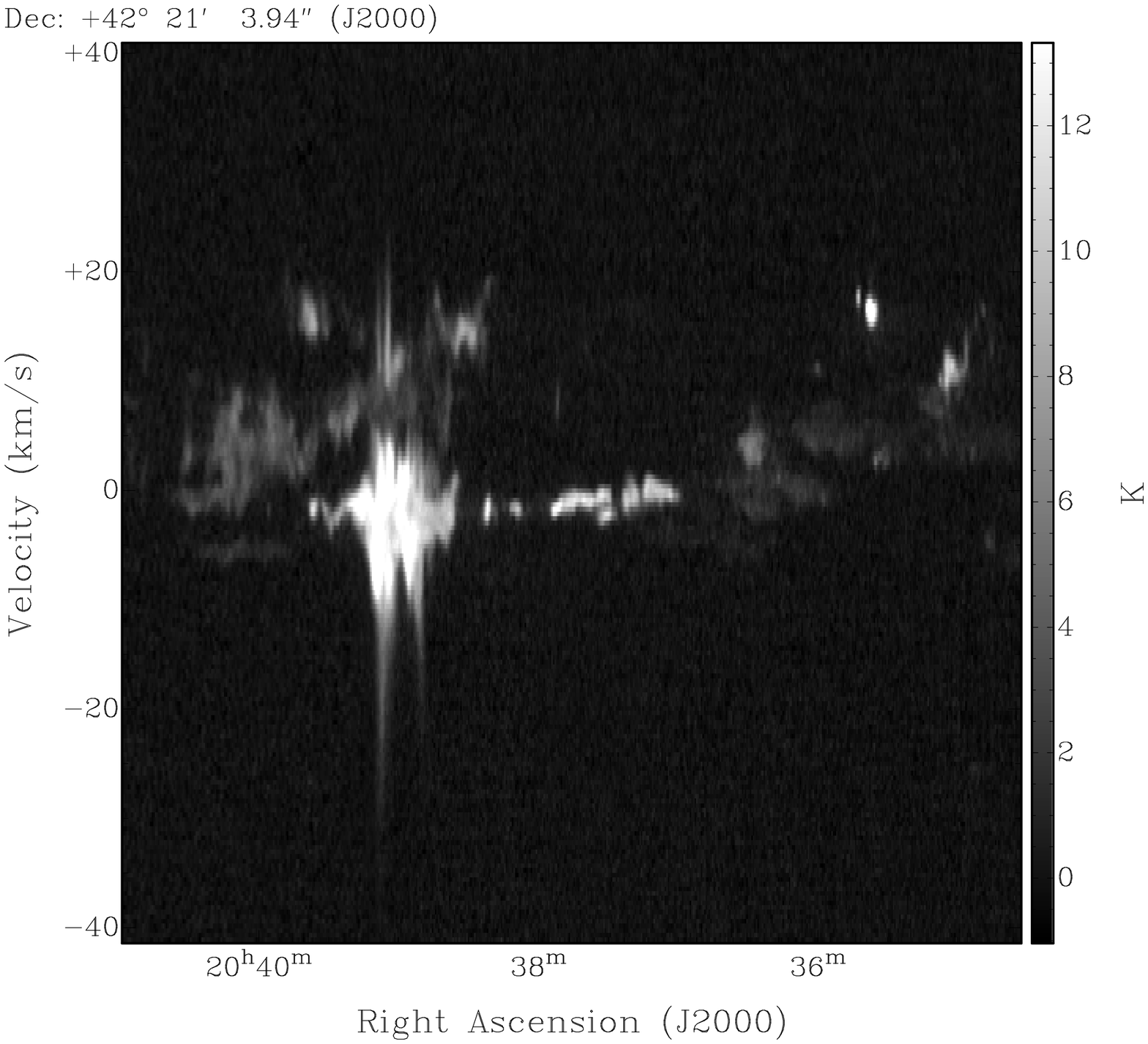}
\end{minipage}
\begin{minipage}{7.5cm}
\includegraphics[bb=25 118 570 725,width=6.4cm,angle=-90,clip]{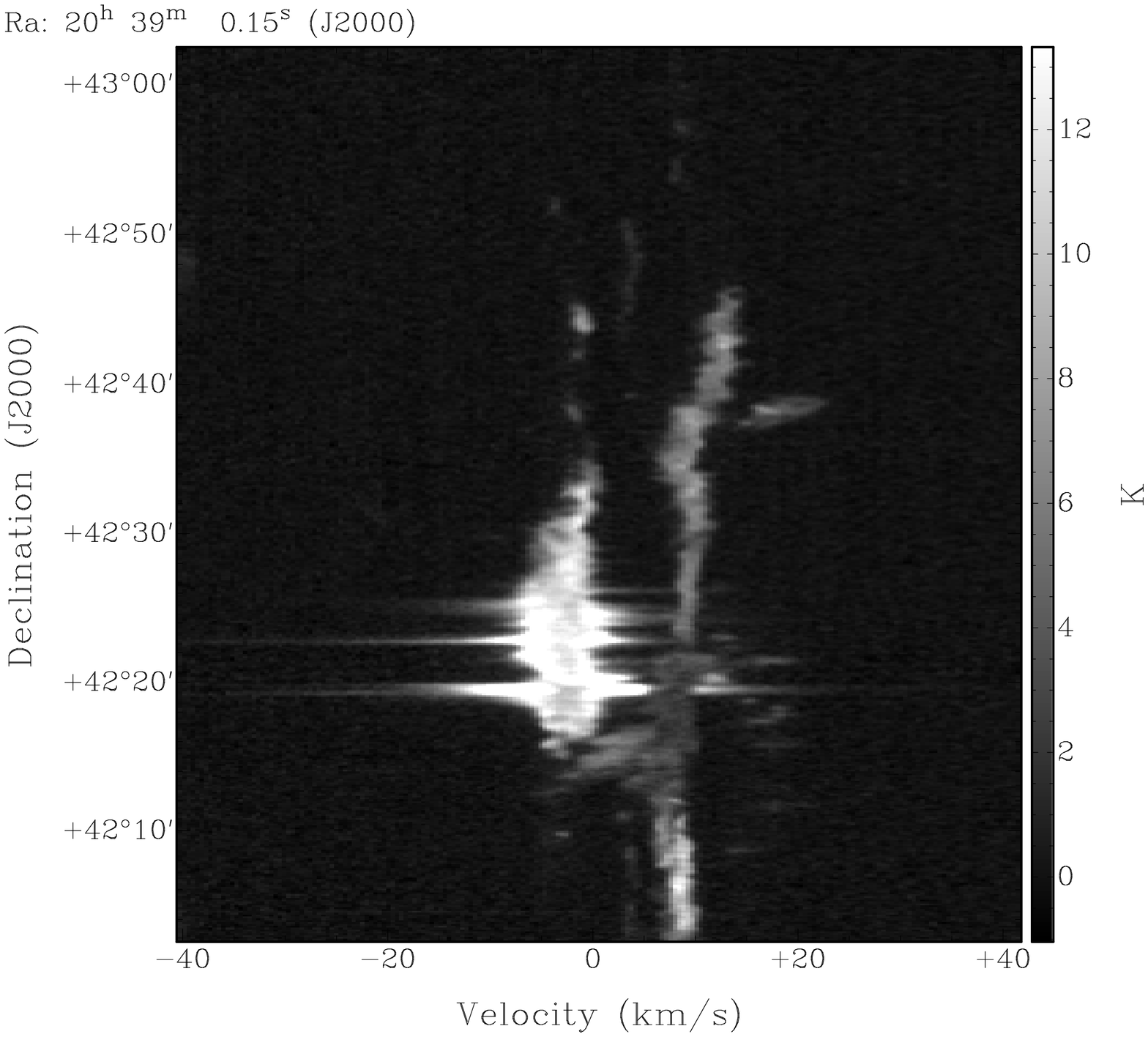}
\end{minipage}
}
\centerline{
\begin{minipage}{7.5cm}
\includegraphics[bb=25 118 570 725,width=6.4cm,angle=-90,clip]{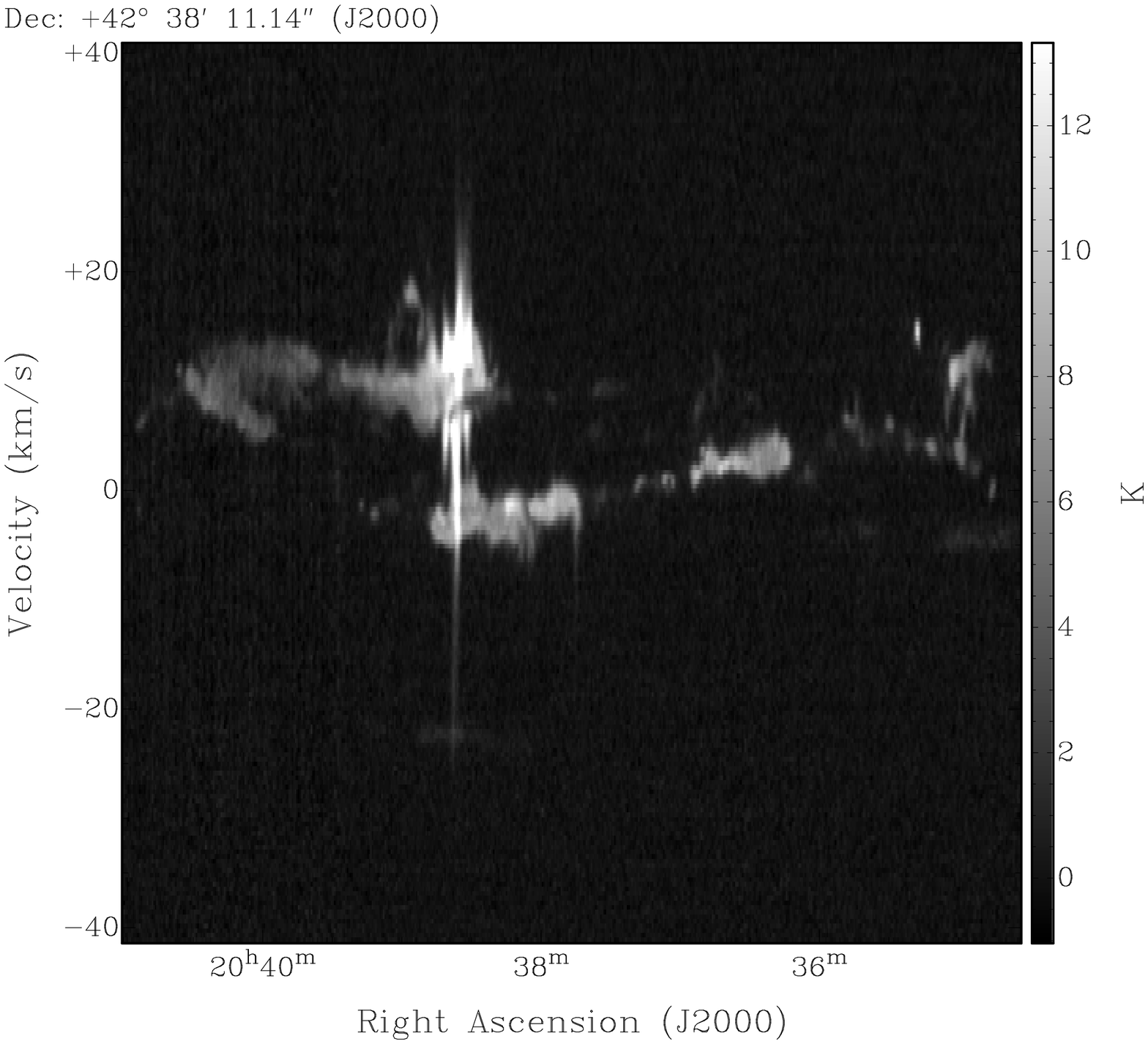}
\end{minipage}
\begin{minipage}{7.5cm}
\includegraphics[bb=25 118 570 725,width=6.4cm,angle=-90,clip]{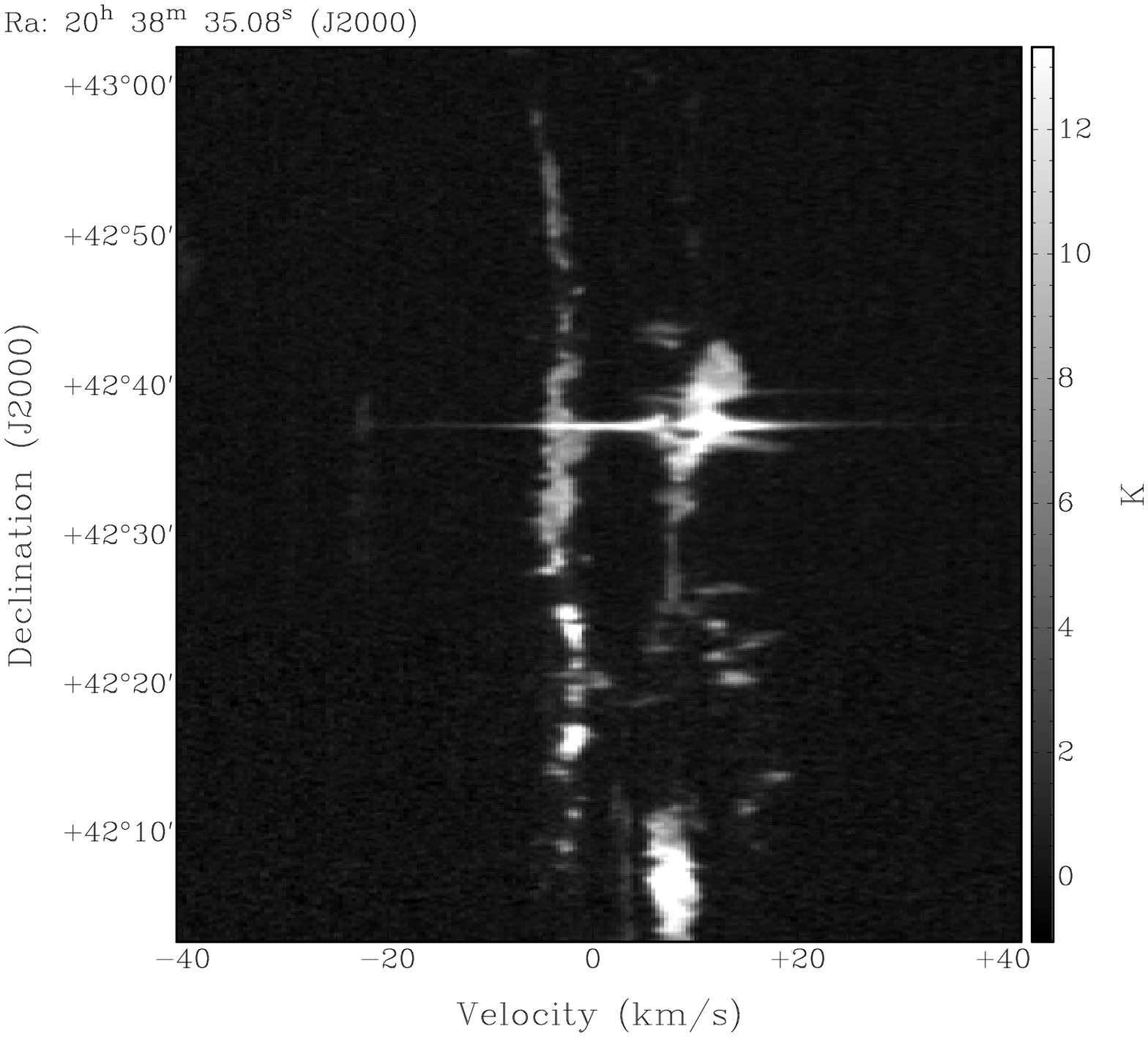}
\end{minipage}
}
\caption{\label{fig:radecvelo} Sample R.A.--v (left) and v--DEC (right)
maps to demonstrate that the apparently very complex and chaotic appearance in the
zeroth and first moment maps shown in Figs.~\ref{fig:fl12} to \ref{fig:fl78} can easily be resolved 
in the third dimension. The two top panels represent an area where all three layers of
\element[][12]{CO}{3--2} emission (see text) that we found in the Pathfinder are
visible, the centre panels were taken towards DR~21, and the bottom panels towards W\,75\,N.}
\end{figure*}

The eastern portion of the Pathfinder is far more complex than the
western region, as can be easily seen by comparing
Figs.~\ref{fig:fl12} to \ref{fig:fl78}. There were major concerns that
the Cygnus~X region might be too crowded and complicated for a project
like ours with many sources located along the line of sight, thus
making physical interpretation of individual objects too difficult
through projection effects and the ravages of ongoing star
formation. Figs.~\ref{fig:fl56} and \ref{fig:fl78} seem at first
glance to confirm that assessment.  The eastern portion of our survey
area was previously studied by \citet{schn06} using the
\element[][12]{CO}(3--2) and \element[][13]{CO}(2--1) \& (3--2) lines
with the KOSMA 3m telescope. Their survey led them to believe that all
the emission in velocity ranges -10~to~+20~km~s$^{-1}$ is dynamically
connected, which would greatly complicate interpretation.

However, in contrast to the study by \citet{schn06}, we detect three
distinct layers of $^{12}$CO(3--2) emission, which do not show any
interconnection. The three layers can be seen clearly separated in the
position-velocity diagrams in Fig.~\ref{fig:radecvelo} (top
panels). We also display RA--v and v--DEC images towards the
well-known compact \ion{H}{ii} regions DR~21 (centre) and W\,75\,N
(bottom) to show that even towards these busy and very dynamic regions
individual features are separable in velocity space. The separation of
the three layers is a direct result of the higher angular resolution
of our data; earlier studies used poorer angular resolution, which
smeared together objects that we now see to be spatially distinct.

Layer 1 in Fig.~\ref{fig:radecvelo} is concentrated in radial
velocities from $\approx -10$~km\,s$^{-1}$ to $\approx
0$~km\,s$^{-1}$. This velocity range contains the \ion{H}{ii} region
complex DR\,21 and other \ion{H}{ii} regions and star formation
regions connected to it.  There is a significant amount of faint
diffuse emission in layer 2 at low positive velocities that moves in
like a curtain of fog, covering everything.  All emission from objects
that have their peak emission in the other velocity ranges, but still
show some emission at low positive velocities (outflows or clouds with
wide emission lines) is completely absorbed by these diffuse clouds.
We believe that these clouds represent the Cygnus Rift, an area of
dark clouds at a distance of 500 to 800~pc, responsible for a jump in
visual extinction for stars observed in this direction (e.g. Schneider
et al., 2006, and references therein). This is the first time that the
molecular line emission of the rift has been separated from the
emission of other molecular clouds in the area.  Layer 3 of
$^{12}$CO(3--2) emission in Fig.~\ref{fig:radecvelo} can be found
above +8~km~s$^{-1}$. This layer encompasses gas primarily connected
with W\,75\,N and related \ion{H}{ii} regions and star formation
complexes.

\citet{maoy09} previously studied the possibility that DR\,21 and
W\,75\,N are two regions in collision. They concluded that the two
regions overlap spatially, but not in velocity and are therefore not
connected.  We find similar results. The outflows from DR\,21 and
W\,75\,N clearly show absorption in their high velocity wings by the
diffuse shroud between +7 and +8~km~s$^{-1}$. Since the velocity
of the emission in the wings is not at all an indication of its
location, and since the bulk of emission from W\,75\,N is at a
different velocity from the bulk of emission from DR\,21, we find no
evidence suggesting the two regions are colliding.

\subsection{Kinematic Distances}
By using a flat rotation curve, with the IAU-supported values for the
Galacto-centric distance of the Sun of $R_\odot = 8.5$~kpc and its
orbital velocity of $\Theta_\odot = 220$~km~s$^{-1}$ we find that
objects should be seen to a maximum radial velocity of around
+4~km~s$^{-1}$. Using other more modern values for $\Theta_\odot$ and
$R_\odot$ lead to similar results. Based upon the new view of the
Milky Way from Spitzer \citep{chur09} the direction we are observing
($\ell\sim 80\degr$) looks through our local spiral arm, the Orion Spur,
to a distance of about 2.5~kpc, through the Perseus arm, and
finally through the Outer Arm. Objects at the maximum velocity of the
rotation curve are those objects closest to the center of our
Galaxy. We clearly find a lot of emission at velocities more positive
than +4~km~s$^{-1}$ with multiple objects seen upwards of
+20~km~s$^{-1}$. This is beyond anything that can be explained by
velocity dispersion.

\citet{burt71} and \citet{robe72} showed that streaming motions caused by a spiral 
density wave pattern can systematically shift the bulk average motion of local gas 
into positive velocities in this part of our Galaxy, since we are located between
major spiral arms, where streaming motions along the orbital direction is at its
maximum. This could account for about $+6$~km\,s$^{-1}$ \citep{wiel79} in our 
observations towards Cygnus~X. Since there is still a high discrepancy in radial velocity the 
most likely scenario for the residual velocity is that these objects are 
being pushed away from us at high speeds, up to about$10$~km\,s$^{-1}$. This
clearly shows that traditional 
kinematic distance estimates in the Pathfinder area cannot be trusted. The main information to be gained from
radial velocities is an object's position and dynamics relative to the large cloud 
complexes present in the area.

\section{Discussion of the Results from the Pathfinder}
		
\begin{figure*}
	\centerline{\includegraphics[bb=40 45 885 585, width=15cm, clip]{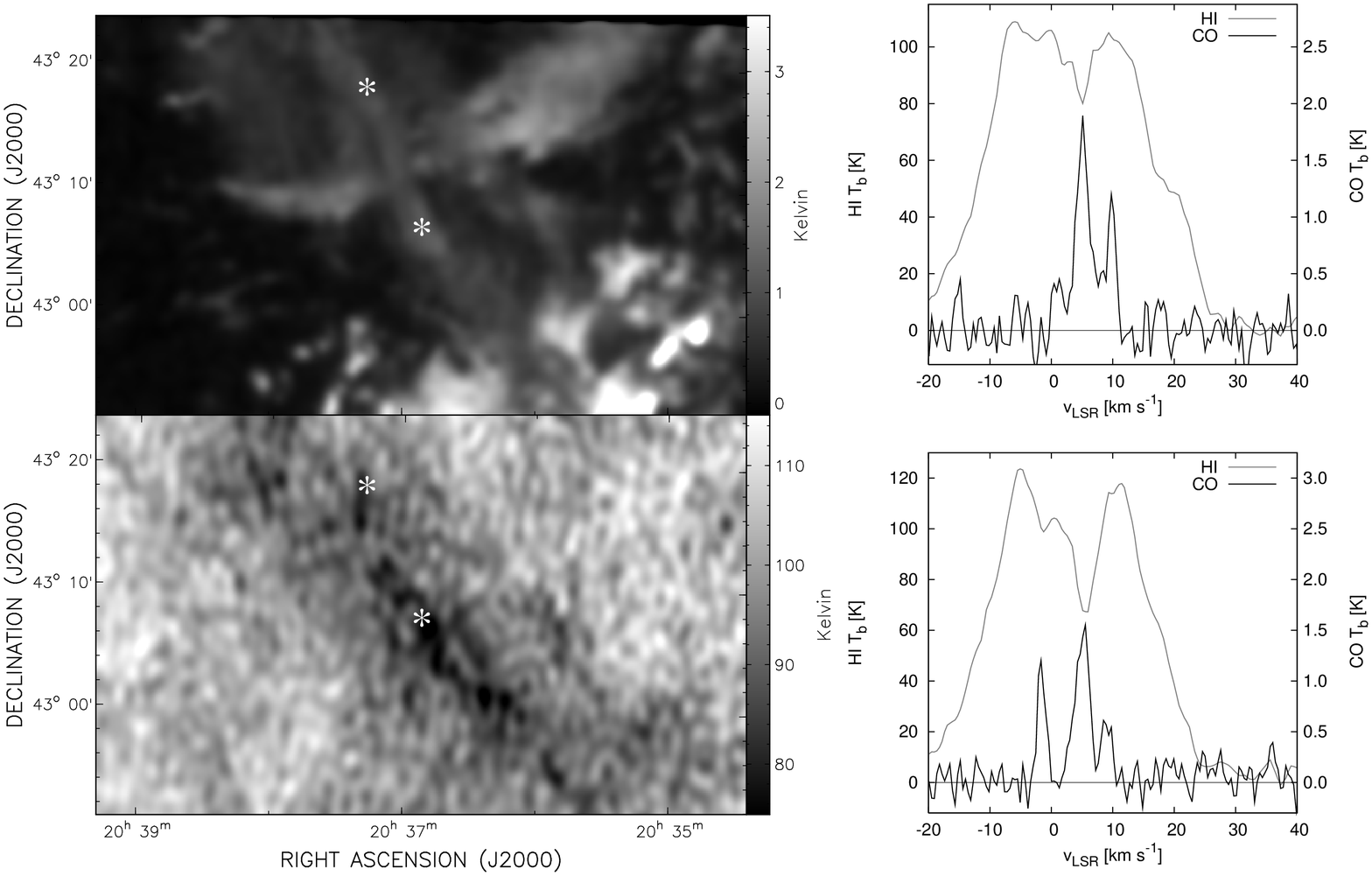}}
	\centerline{\includegraphics[bb=25 55 850 545, width=15cm, clip]{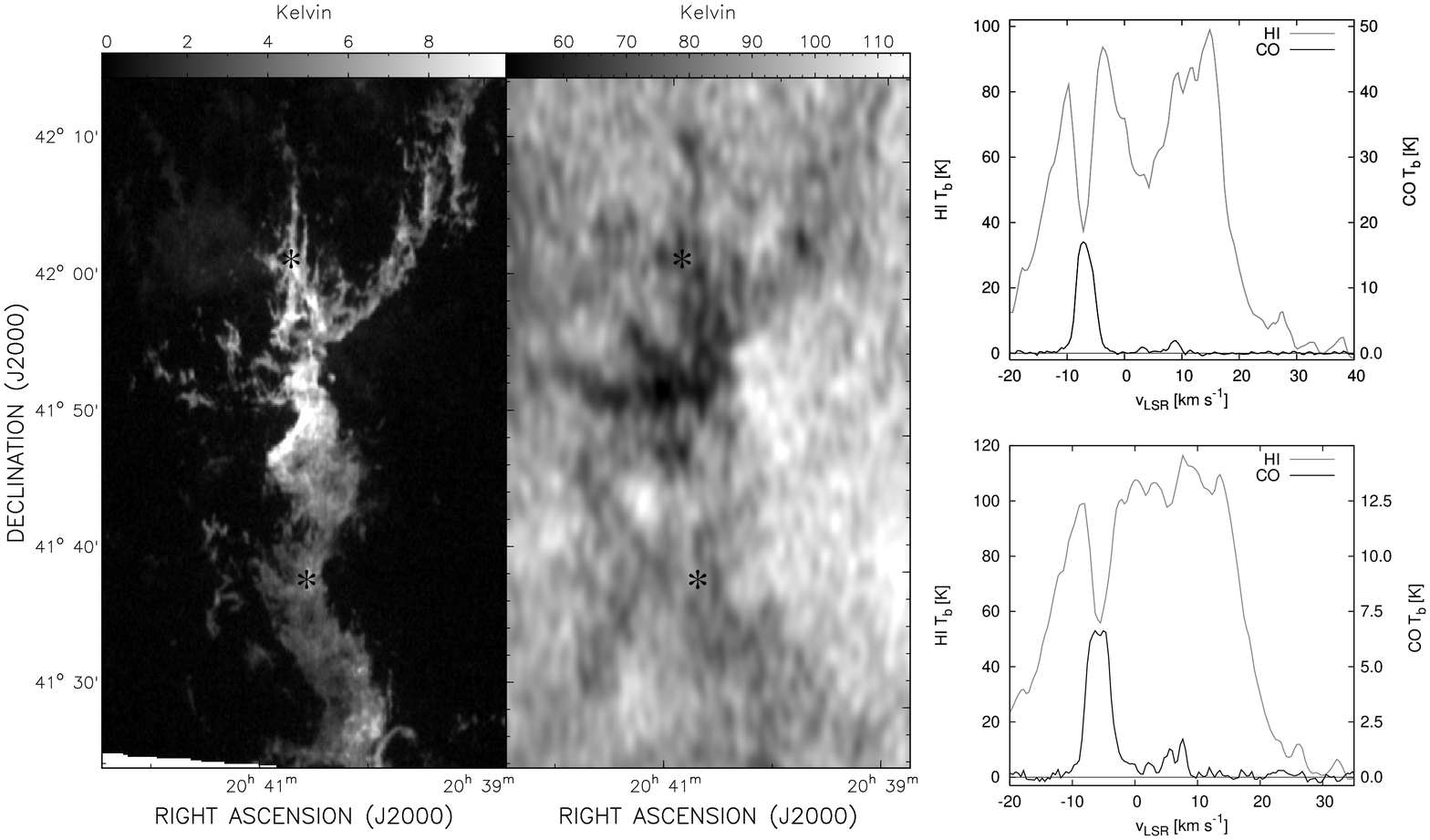}}
	\caption{Two examples of clouds of $^{12}$CO(3--2) emission that coincide with
	extended features of \ion{H}{i} self absorption. In the left $^{12}$CO(3--2) 
	emission maps are shown with corresponding \ion{H}{i} images. On the right CO and \ion{H}{i}
	profiles are displayed. These were taken at the positions marked with the asterisks 
	in the images. In the top we show an example for a Cygnus Rift cloud at a velocity
	of $+5$~km\,s$^{-1}$. The bottom represents a cloud at low negative velocities
	connected to DR~21. The images represent a velocity of $-8$~km\,s$^{-1}$.}
	\label{fig:hisa}
\end{figure*}

In this section we make a brief analysis of the Pathfinder data according to the
different scientific objectives laid out in Section 2.

\subsection{Cold Molecular Clouds, HISA, and the ``Great Cygnus Rift''}

\citet{gibs05} investigated \ion{H}{i} self absorption features in the Galactic plane in the Outer
Galaxy using data from the Canadian Galactic Plane Survey \citep{tayl03}. In their
catalogue of major CGPS HISA features \citep[Table~1]{gibs05} all HISA in Cygnus~X
is listed as one major feature GHISA 079.88+0.62+02. This is by far
the largest HISA feature in this catalogue, more than 4 times bigger than the second largest. It
also shows the largest velocity coverage from about $-10$~km\,s$^{-1}$ to 
$+15$~km\,s$^{-1}$. According to \citet{gibs05} most stronger HISA is 
organized into discrete complexes, many of 
which follow a longitude-velocity distribution that suggests that they have been made 
visible by the velocity reversal of the Perseus arm's spiral density wave. The cold
\ion{H}{i} revealed in this way may have recently passed through the spiral shock and be on 
its way to forming molecules and, eventually, new stars. 

In the area of the Cygnus~X Pathfinder we find HISA in two of the three previously described
layers of $^{12}$CO(3--2) emission from $\approx -8$~km\,s$^{-1}$
to $\approx +10$~km\,s$^{-1}$. In layer 3 no significant
HISA feature is detected. Layer 2 shows most of the significant
HISA, which can be found in almost the entire Pathfinder area. All of these features seem
to be related to large and diffuse $^{12}$CO(3--2) emission clouds of very low surface brightness.
These clouds must be tremendously dense and cold, because they have, in addition to
the low surface brightness, very small
line width and completely absorb all the emission from background clouds that reaches into their
velocity range. One example is shown in the upper panel of Fig.~\ref{fig:hisa}. 

Layer 1
also shows a lot of complex HISA, in particular in the eastern area, but also in the upper part of the
western region. But most of the HISA in this velocity range does not correspond to any $^{12}$CO(3--2) emission
in the Pathfinder. In Fig.~\ref{fig:hisa} (lower panels) we display an example of a very bright molecular cloud connected
to the DR~21 complex that does
correspond to a HISA complex with very deep absorption.

Before analysing the detected HISA features and their corresponding
molecular clouds we need to discuss briefly under what
circumstances HISA is seen. The most important characteristic of
HISA is that two separate \ion{H}{i} clouds along the line of sight
with overlapping radial velocity are needed. The background cloud must
be a warm \ion{H}{i} cloud of high radio surface brightness while the
foreground cloud -- the absorbing cloud -- must be cold and of high
optical depth implying a high density. In addition the amplitude of
the absorbed signal must be higher than the emission of the absorbing
cloud. In the Canadian Galactic Plane Survey most of the HISA features
are found in the Perseus arm where the cold foreground gas is located
in the spiral shock and the warm background is interarm gas from
beyond the Perseus arm \citep{gibs05}.

Another typical location for HISA would be the inner Galaxy where the
differential Galactic rotation produces a distance ambiguity for any
given radial velocity from our point of view. Towards the Cygnus~X
region the line of sight passes through the inner Galaxy and returns
to the Sun's galactocentric radius at a distance of about 3.0~kpc,
assuming the IAU endorsed value for $R_\odot$. This is much farther
than any significant object in the Cygnus~X region is believed to
be. This implies that all molecular clouds in Cygnus~X should be found
within a radial velocity range of $\approx 0$~km\,s$^{-1}$ to $\approx
+10$~km\,s$^{-1}$ based entirely on their orbital motion around the
Galactic centre (see discussion in Section~4.3). Any clouds with
radial velocities outside that range must have been either pushed away
from or towards us by stellar winds or supernova shock waves.

Under these circumstances it is not surprising that we do not find
HISA corresponding to the $^{12}$CO(3--2) emission layer 3. In
particular radial velocities in excess of $+20$~km\,s$^{-1}$ are
highly forbidden in view of Galactic rotation models. The only way
that clouds can have received such a high positive radial velocity is
that some force has pushed them away from us. It would be very
unlikely that this has happened twice at separate distances
along the line of sight with a warm cloud being pushed away with just
exactly the velocity needed to be the background for a cold cloud that
is pushed away from us too. This does not mean that there are no cold
dense \ion{H}{i} clouds present. We just do not have the bright
background to reveal them in self absorption.

The HISA features related to low surface brightness molecular clouds
in layer 2 are a typical result of the inner Galaxy distance ambiguity
and can be easily explained by a flat rotation curve combined with
density wave streaming motion (see section 4.3). In this case the cold
and dense absorbing clouds must be located between the Sun and the
tangent point at about 1.5~kpc (the location closest to the Galactic
center) and the warm background gas must have a distance between 1.5
and 3.0~kpc. The most likely explanation for the large amount of cold
atomic and molecular gas together is that this layer of
$^{12}$CO(3--2) emission represents the Great Cygnus Rift, a large
molecular cloud and dust complex that reveals itself by a large jump
in observed visual extinction of stars at a distance of about 500 to
800~pc \citep[][and references therein]{schn06}. There is a lot of
evidence to support this explanation. The molecular gas and HISA
observed in this velocity range can be found almost everywhere over
the mapped region and the observed radial velocity indicates a
location closer than 1.5~kpc. The atomic and molecular gas must be
very cold and very dense, as evidenced by the deep absorption signals
these clouds cause. These same clouds seem to be related to smooth
dust emission seen in the $8\mu$ maps of the MSX survey (although the
Cygnus~X area is so complex that a positive identification cannot be
made). This likely combination of intermixed cold dust, atomic and
molecular gas represents the perfect environment for the formation of
molecules. This would be the first time that the molecular gas
related to the Cygnus Rift has been separated from other clouds in
velocity space.

The HISA features related to layer 1 are not as easily
explained. A flat rotation curve combined with density wave streaming motion does not account for a distance ambiguity
for gas at negative velocities in the direction of Cygnus~X. This can be explained by assuming that the warm background \ion{H}{i} clouds are
the ones predicted by Galactic rotation models, which would locate them at a distance of about 3 to 4~kpc, an
interarm location between the local arm and Perseus arm. The foreground cold and dense absorbing clouds must have
been pushed towards us by an external force like stellar winds or supernova shock waves. This would imply
a distance smaller than 3~kpc.

It is interesting to note, that within the Cygnus Rift, literally every HISA feature is accompanied by a detectable
molecular cloud, while most of the HISA at low negative velocities does not correlate with $^{12}$CO(3--2) emission.
In the Canadian Galactic Plane Survey HISA clouds often have no CO counterparts, and \citet{gibs05} conclude that
these clouds trace atomic gas on the path to formation of molecular hydrogen where CO molecules have not yet 
formed. This could indicate that most of the dense clouds at low negative velocities are in an earlier phase
of the molecule formation process, while the clouds of the Cygnus Rift are further developed. However, a 
contrary indication would be the fact that in layer 1 more star formation, as revealed by molecular outflows,
is found than in
the Cygnus Rift, indicating that the Cygnus Rift clouds had less time yet to develop gravitational 
instabilities and collapse. On the other hand, outflows naturally are only found to be connected to HISA,
which are related to observable molecular clouds. The Cygnus Rift may represent just a snapshot in the 
evolutionary path of molecule formation, while layer 1 represents
a large spread along that evolutionary path. Clearly observations over a wider area are required to make a more
thorough investigation of the evolutionary state of the molecule formation process in Cygnus~X.

\subsection{Molecular Outflows}

\begin{table*}
\caption{\label{tab:of} A list of outflows detected in the Pathfinder. Displayed are an identifier equatorial
coordinates radial velocity and a possibly related infrared source.}
\centerline{
\begin{tabular}{l c c c l}
Identifier & $\alpha$ [$^\mathrm{h~m~s}$] & $\delta$
[$^{\circ}$~$^{\prime}$~$^{\prime\prime}$] & $v_{\mathrm{lsr}}$ [km~s$^{-1}$] &
Infrared Source \\
\hline
G79.886+2.552 & 20 24 31.6 & 42 04 20 & 6.6 & IRAS 20227+4154   \\
G80.158+2.727 & 20 24 35.7 & 42 23 41 & 6.0 & IRAS 20228+4215   \\
G80.149+2.710 & 20 24 38.6 & 42 22 42 & 6.0 & 2MASS J20243430+4221265   \\
G79.962+2.556 & 20 24 44.7 & 42 08 11 & 5.5 & IRAS 20231+4157   \\
G81.424+2.140 & 20 31 12.3 & 43 04 53 & -1.7 & IRAS 20294+4255  \\
G80.314+1.330 & 20 31 12.3 & 41 42 30 & -31.4 & 2MASS J20311521+4142249 \\
G81.435+2.147 & 20 31 12.5 & 43 05 42 & -2.4 & IRAS 20294+4255  \\
G81.302+1.055 & 20 35 33.5 & 42 20 17 & 14.6 & 2MASS J20353805+4220318 \\ 
G80.815+0.661 & 20 35 40.9 & 41 42 44 & -3.0 & 2MASS J20352935+4143345  \\
G80.972+0.700 & 20 36 01.2 & 41 51 41 & 6.0 & IRAS 20342+4139   \\
G81.218+0.877 & 20 36 03.3 & 42 09 50 & 13.6 &  \\
G80.832+0.570 & 20 36 07.6 & 41 40 15 & 11.5 & IRAS 20343+4129  \\
G81.140+0.687 & 20 36 37.2 & 41 59 16 & 12.3 &  \\
G81.539+0.983 & 20 36 38.5 & 42 29 02 & 2.6 &   \\
G81.829+1.195 & 20 36 40.2 & 42 50 35 & 13.6 &  \\
G80.866+0.415 & 20 36 53.6 & 41 36 17 & -4.7 & IRAS 20350+4126  \\
G81.365+0.781 & 20 36 56.7 & 42 13 26 & 16.6 &  \\
G81.340+0.755 & 20 36 58.6 & 42 11 17 & 15.7 & 2MASS J20365781+4211303  \\ 
G80.862+0.385 & 20 37 00.6 & 41 35 00 & -1.4 & IRAS 20352+4124  \\
G81.437+0.728 & 20 37 24.5 & 42 14 57 & -1.3 & MSX6C G081.4515+00.7280  \\
G80.916+0.331 & 20 37 25.0 & 41 35 39 & -1.3 & IRAS 20355+4124  \\
G81.460+0.735 & 20 37 27.3 & 42 16 19 & -1.3 & 2MASS J20372634+4215518  \\
G81.462+0.736 & 20 37 27.3 & 42 16 27 & 17.4 &  \\ 
G81.794+0.910 & 20 37 47.6 & 42 38 37 & -0.4 & 2MASS J20374746+4238370  \\
G81.861+0.959 & 20 37 48.0 & 42 43 35 & -0.8 &  \\
G81.813+0.919 & 20 37 49.1 & 42 39 51 & -0.8 & 2MASS J20374733+4240529  \\
G81.770+0.851 & 20 37 58.3 & 42 35 19 & -0.4 &  \\
G81.857+0.905 & 20 38 01.4 & 42 41 25 & -3.0 & 2MASS J20375536+4240468          \\
G81.831+0.880 & 20 38 02.6 & 42 39 17 & -2.1 & 2MASS J20380710+4238521  \\
G81.634+0.726 & 20 38 03.7 & 42 24 18 & 8.5 &   \\
G81.867+0.780 & 20 38 35.9 & 42 37 22 & 10.6 & W75N     \\
G81.773+0.687 & 20 38 41.3 & 42 29 28 & -3.8 & [CGC93] FIR 11   \\
G81.681+0.540 & 20 39 01.1 & 42 19 43 & -3.0 & DR21     \\
G81.663+0.468 & 20 39 15.9 & 42 16 15 & 19.9 & 2MASS J20391672+4216090  \\
G81.117-0.140 & 20 40 04.1 & 41 28 03 & -4.2 & MSX6C G081.1225-00.1343  \\
G81.175-0.100 & 20 40 05.1 & 41 32 15 & -3.8 &  \\
G81.559+0.183 & 20 40 08.6 & 42 00 52 & 9.3 &   \\
G81.551+0.098 & 20 40 28.7 & 41 57 21 & -5.9 &  \\
G81.450+0.016 & 20 40 29.7 & 41 49 35 & -4.7 &  \\
G81.582+0.104 & 20 40 33.3 & 41 59 05 & -5.9 & 2MASS J20403814+4200055 \\ 
G81.476+0.020 & 20 40 33.8 & 41 50 56 & -4.7 &  \\
G81.317-0.103 & 20 40 33.9 & 41 38 53 & -5.1 & HFE 69   \\
G81.632+0.102 & 20 40 43.7 & 42 01 21 & -6.8 &  \\
G82.186+0.105 & 20 42 33.1 & 42 27 42 & 9.8 &   \\
G82.189-0.042 & 20 43 11.6 & 42 22 24 & 7.2 &   \\
G82.581+0.203 & 20 43 27.8 & 42 49 58 & 11.1 & MSX6C G082.5828+00.2014  \\
G82.571+0.194 & 20 43 27.9 & 42 49 11 & 11.1 &  \\
\hline
\end{tabular}}
\end{table*}

The presence of a bipolar outflow, together with a dusty disk and envelope, signifies the presence 
of a protostar. The outflow mechanism is believed to be an
energetic mass-ejection phenomenon associated with mass accretion in the very early stages 
of stellar evolution. It is most energetic at the earliest Class~0 and 1 phases of proto-stellar development.
The high critical density of
$^{12}$CO(3-2) makes it an excellent tracer of dense material and
line observations with the JCMT are probably the most
sensitive method of detecting the warm outflow gas \citep{dent09}. Previous outflow
searches have generally been biased toward near-infrared signposts of star
formation. It is therefore of vital importance to make a complete and unbiased
search for molecular outflows in a large molecular cloud and
star formation complex such as the Cygnus X region. 

The outflowing mass initially leaves the protostar with speeds several times the escape speed
(several $\times 100$~km\,s$^{-1}$), 
but collides with the local parent cloud slowing the jet down 
and creating shocks and gas velocities of a few $\times 10$~km\,s$^{-1}$ \citep{stah05}. 
Depending 
on the density and thickness of the parent cloud an outflow will pierce through the local ISM 
or it will be slowed down eventually to the velocity of the local ISM. By examining the 
spectra of the outflow a basic understanding of the outflow's history can be ascertained. This 
is done by studying the evolution of the outflow over velocity. 

We searched the whole data cube by eye, examining it in
RA~$\times$~v and v~$\times$~DEC slices. High velocity wings are
readily seen stretching away from the core along the velocity
axis. Examples can be seen in Fig.~\ref{fig:radecvelo} (lower two
panels). We include any source with a spectrum containing a width at
its base of over 8~km~s$^{-1}$ that is clearly non-Gaussian in
shape. Towards a few of the outflows other objects can be seen
contaminating the spectra (i.e. they appear to lengthen the wings, or
cause another peak off the central velocity), so a careful analysis of
the cube in RA~$\times$~DEC~$\times$~v needs to be carried out.
This contamination often arises from extended emission that
covers the area, or from small dense clouds overlapping
spatially with some part of the outflow.
		
\begin{figure*}
	\centerline{\includegraphics[bb=12 70 805 445, width=17cm,clip]{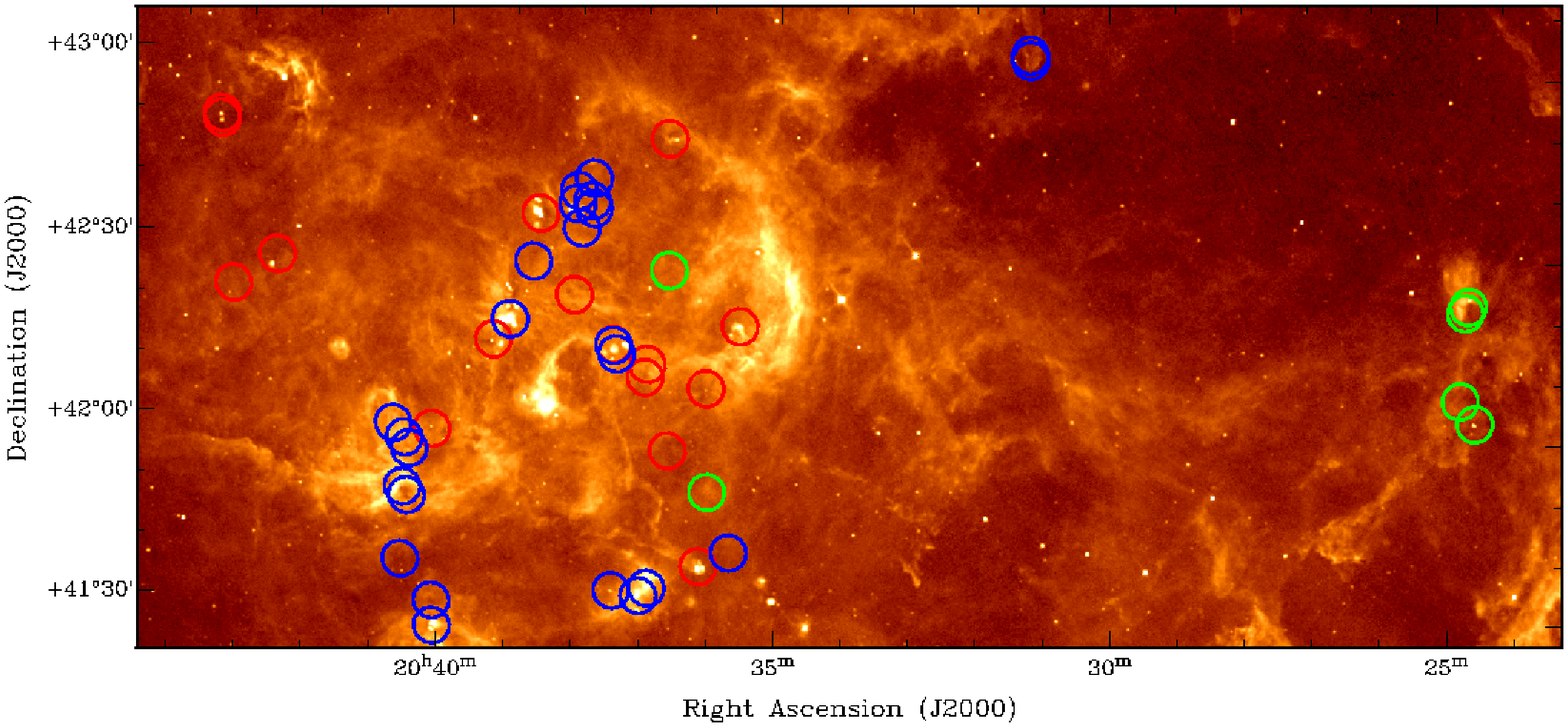}}
	\centerline{\includegraphics[bb=12 55 805 460, width=17cm,clip]{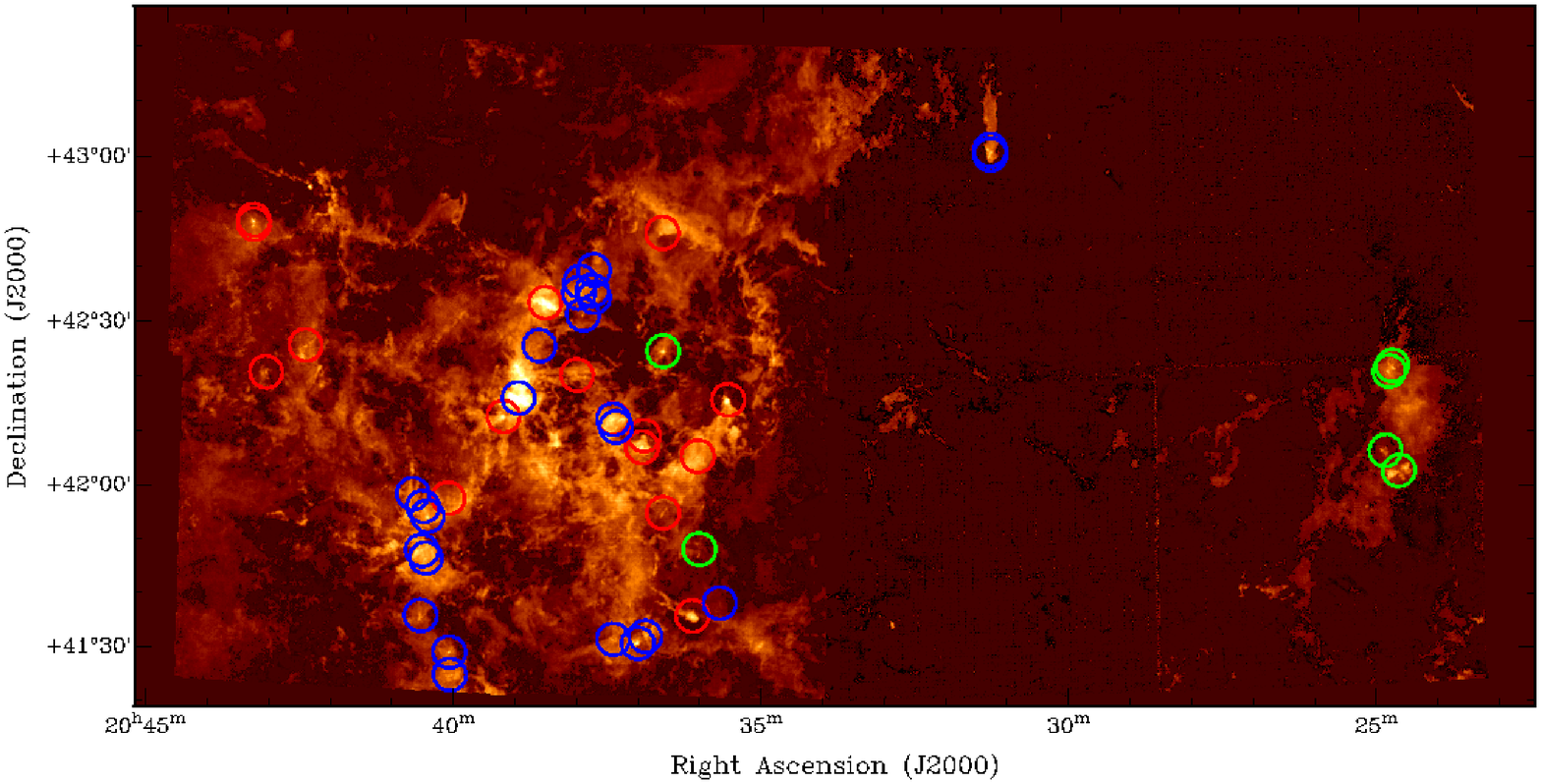}}

	\caption{\label{fig:ofall}MSX image at $8\mu$ (top) and an
	integrated $^{12}$CO(3--2) image (bottom), the zeroth moment
	map between $-20$~km\,s$^{-1}$ and $+30$~km\,s$^{-1}$ of the
	Pathfinder area. The location of all 47 outflows that we
	detected are indicated. Blue circles indicate a negative radial
	velocity for the parent cloud of the outflow, green circles a
	connection with the Cygnus Rift (0~km\,s$^{-1} \le v_{rad} \le
	8$~km\,s$^{-1}$, and outflows with velocities above
	8~km\,s$^{-1}$ for their parent cloud are marked by red
	circles.}

\end{figure*}
		
\begin{figure*}
 \centerline{
   \begin{minipage}{6.8cm}
	\includegraphics[bb=50 40 455 760, width=6.8cm,clip]{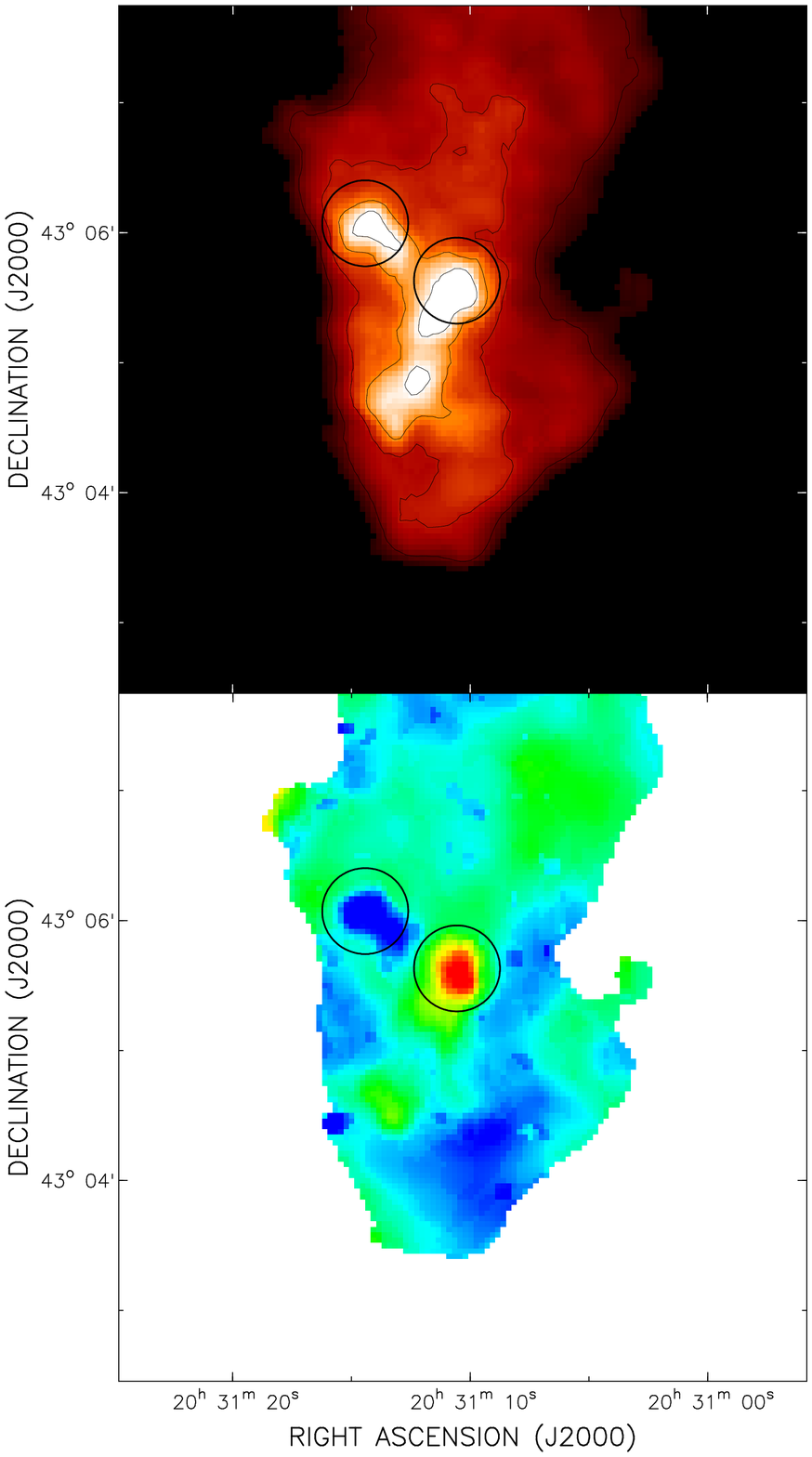}
   \end{minipage}
   \begin{minipage}{6.8cm}
	\includegraphics[bb=50 40 455 760, width=6.8cm,clip]{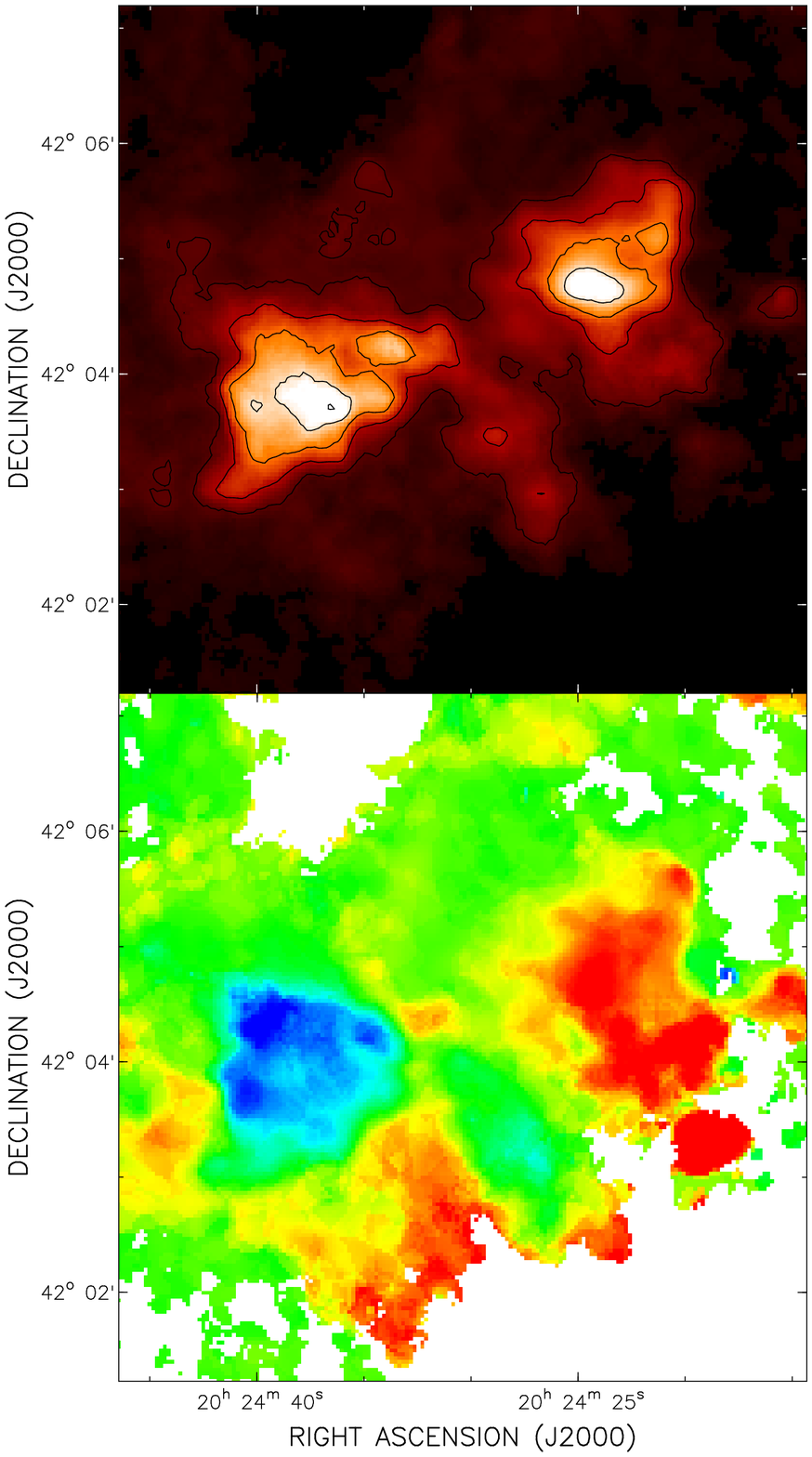}
   \end{minipage}}
	\caption{\label{fig:ofsample} Two examples of molecular outflows detected in the $^{12}$CO(3--2) Pathfinder.
	Zeroth moment maps are displayed in the top and first moment maps in the bottom. On the
	left we show the newly discovered outflow connected to IRAS\,20294+4255 (black circles indicate the locations
	of the blue- and red-shifted lobes). The first
	moment map is displayed in the velocity range from $-5$~km\,s$^{-1}$ (blue) to 
	$-0.5$~km\,s$^{-1}$ (red). On the right we show the large massive outflow connected
	to IRAS~20227+4154. The first moment map is displayed in the velocity range from
	$+1.5$~km\,s$^{-1}$ (blue) to $+8.5$~km\,s$^{-1}$ (red).}
\end{figure*}
		
\begin{figure*}
 \centerline{
   \begin{minipage}{7cm}
	\includegraphics[bb=50 45 540 505, width=7cm,clip]{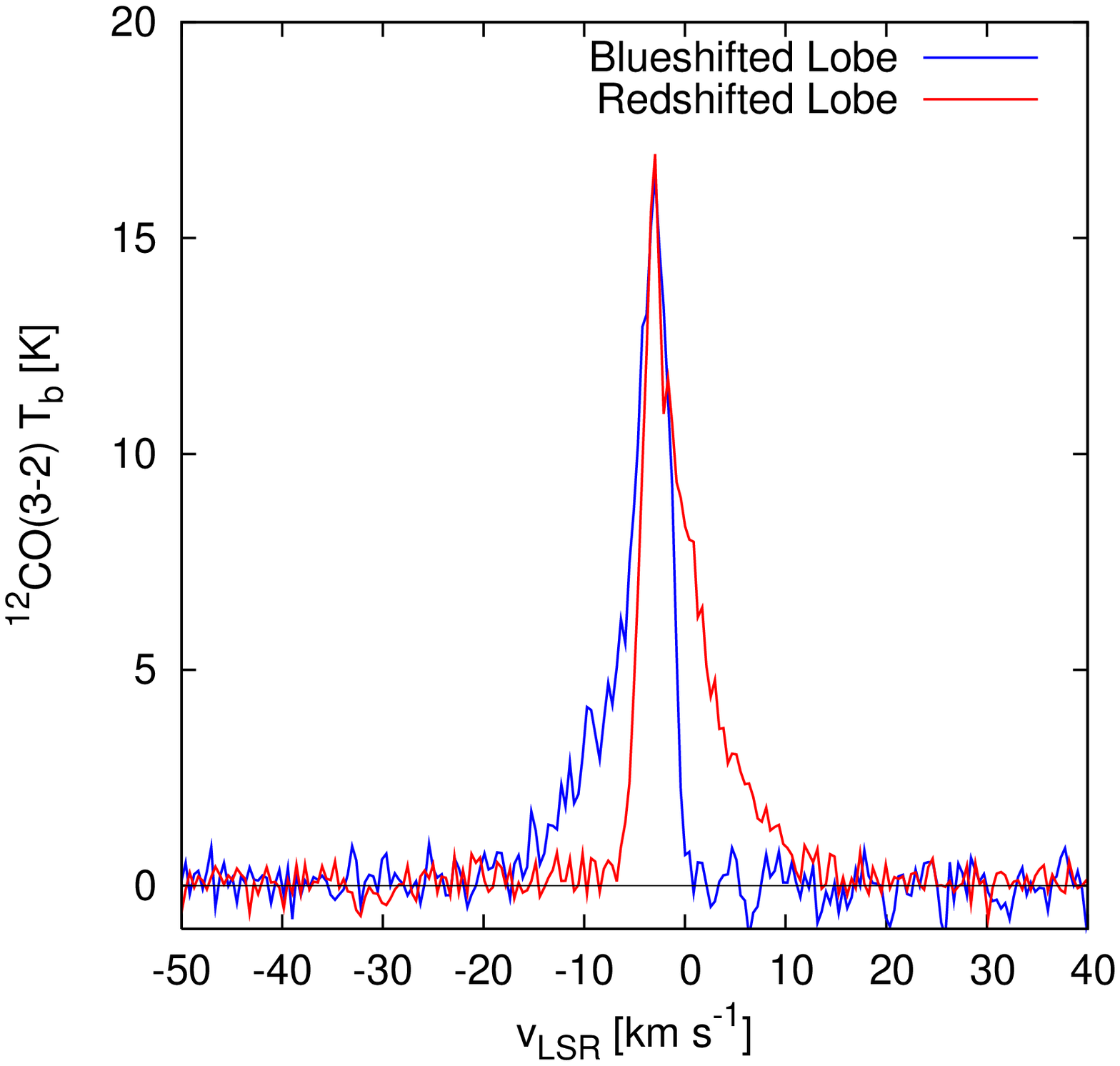}
   \end{minipage}
   \begin{minipage}{7cm}
	\includegraphics[bb=50 45 540 505, width=7cm,clip]{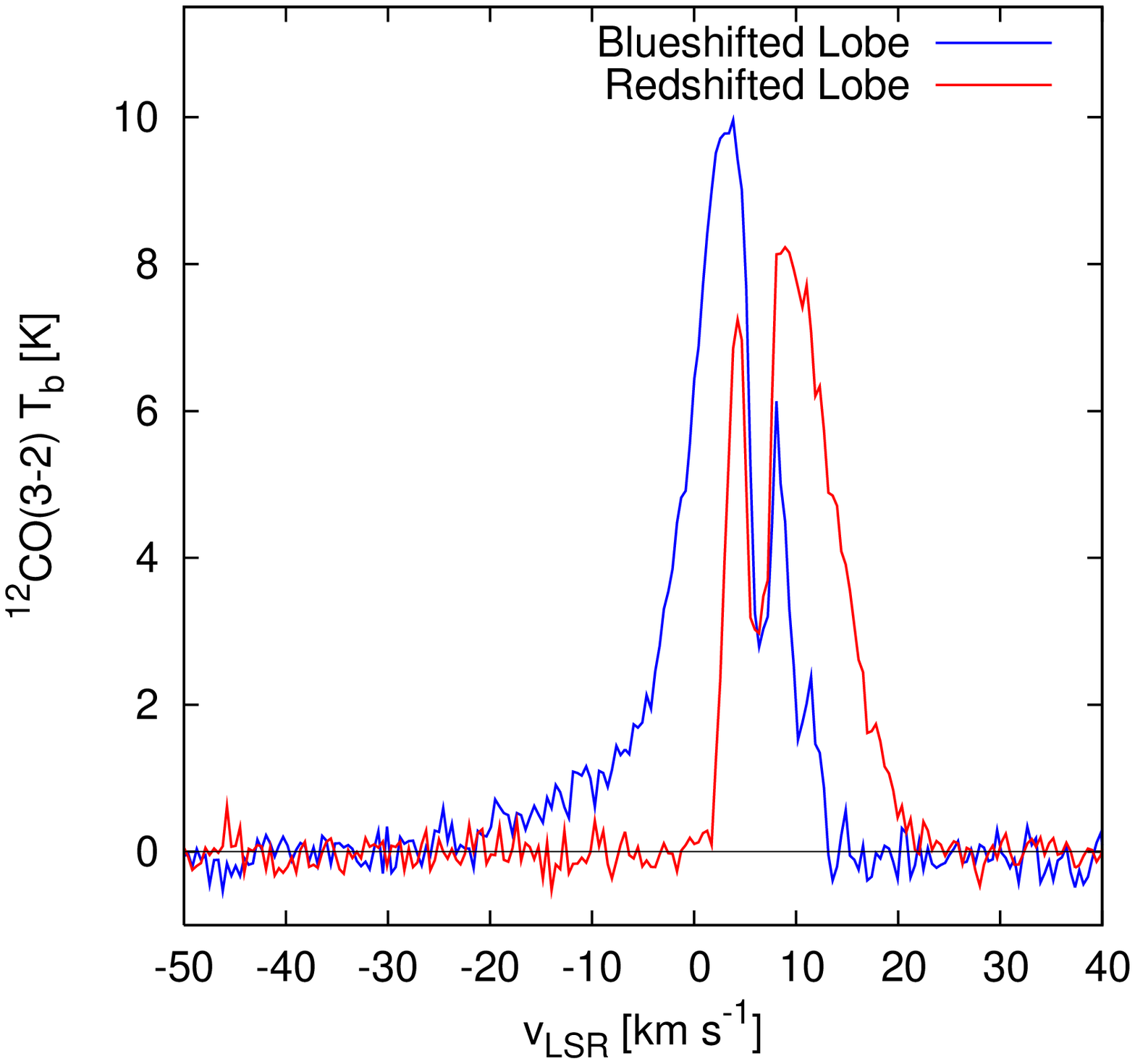}
   \end{minipage}}
	\caption{Two examples of $^{12}$CO(3--2) emission profiles towards the emission peaks of the blue-shifted and 
	red-shifted lobes of the outflows connected to IRAS\,20294+4255 (left) and 
	IRAS~20227+4154 (right).}
	\label{fig:ofprof}
\end{figure*}

We found 47 outflows (listed in Table~\ref{tab:of}) in the
$^{12}$CO(3--2) data cube of the Pathfinder with 27 of them being new
discoveries. The very complex multi-outflow cores related to DR~21 and
W\,75\,N are counted as one outflow each in our list. 46 of these
outflows fall within the velocity range of the Cygnus~X region and of
them -- G80.314+1.330 -- is likely a Perseus arm object with a radial
velocity of about $-31$~km\,s$^{-1}$. To the best of our knowledge,
within our observed survey area we have detected all previously
confirmed molecular outflows that were detected in CO observations.


Most of the molecular outflows identified in this Pathfinder project
have an associated IRAS point source to pin-point the location of the
young stellar object, but 18 of them do not (see
Table~\ref{tab:of}). Owing to the complex nature of this region
continuum surveys such as IRAS pick up everything along the line of
sight making it hard to separate objects. Many of these outflows
discovered in \element[][12]{CO}{(3--2)} were found to overlap
spatially with bright radio sources such as DR\,17 and DR\,23, but at
very different velocities. Since IRAS is a continuum survey and has
much poorer resolution than our \element[][12]{CO}{(3--2)} data many
IR sources that could be associated with these new outflows are lost
through confusion with brighter more extended objects.

In Fig.~\ref{fig:ofall} we display all outflows on an MSX image
of the Pathfinder area and on a zeroth moment map of the Pathfinder
\element[][12]{CO}{(3--2)} survey over the velocity range from
$-20$~km\,s$^{-1}$ to $+30$~km\,s$^{-1}$.  Outflow markers are colour
coded to indicate their location within one of the three
\element[][12]{CO}{(3--2)} emission layers that we have
identified in velocity space. There are only very few outflows at
low positive velocities except for those likely related to the
\ion{H}{ii} region complex NGC~6914. This indicates that there is
little star formation related to the Cygnus Rift in the area of the
Pathfinder. Most outflows display negative velocities and are related
to molecular clouds of bright \element[][12]{CO}{(3--2)} emission that
are dynamically connected to DR~21. The only exception is the
previously mentioned cometary feature related to IRAS~20294+4255,
which contains two molecular outflows. The outflows with a high
positive radial velocity seem to be mostly related to DR~17, Cl~12,
and Cl~14 (see section 5.3).

As examples we display zeroth and first moment maps of two outflows
related to IRAS~20294+4255 and IRAS~20227+4154 and their spectra in
Figs.\ref{fig:ofsample} and \ref{fig:ofprof}. The outflow lobes reveal
themselves in the zeroth moment map as extended emission features,
which have blue- or red-shifted velocities in their first moment
maps. The blue- and red-shifted velocity wings are obvious in the
emission spectra. The spectra of the bipolar outflow related to
IRAS~20227+4154 also show a deep absorption line at about
$+6$~km\,s$^{-1}$ in both the red and blue-shifted spectra. This
indicates that the outflow with its proto-star is not related to the
large diffuse cloud overlapping with it, but must be located behind
it.

A comparison of the optically thick \element[][12]{CO} lines, $J$=3--2
\& $J$=2--1, to the optically thin \element[][13]{CO} lines, $J$=3--2
\& $J$=2--1, was carried out by \citet{choi93} to investigate the
optical depth in the line wings of molecular outflows. For the
$J$=3--2 lines they found that the optical depth decreases further
into the line wings. The lowest velocity has the highest optical depth
and physical parameters are underestimated when only the
optically thick $^{12}$CO line is used.  This can be remedied with
observations of the outflows in an optically thin line. A thorough
study of a number of outflows detected in the Pathfinder based on new
JCMT observations of the more optically thin $^{13}$CO(3--2) and
C$^{18}$O(3--2) lines is being prepared and will soon be ready for
publication.

\subsection{Triggered Star Formation}
		
\begin{figure}
	\centerline{\includegraphics[bb=45 35 640 660, width=9cm,clip]{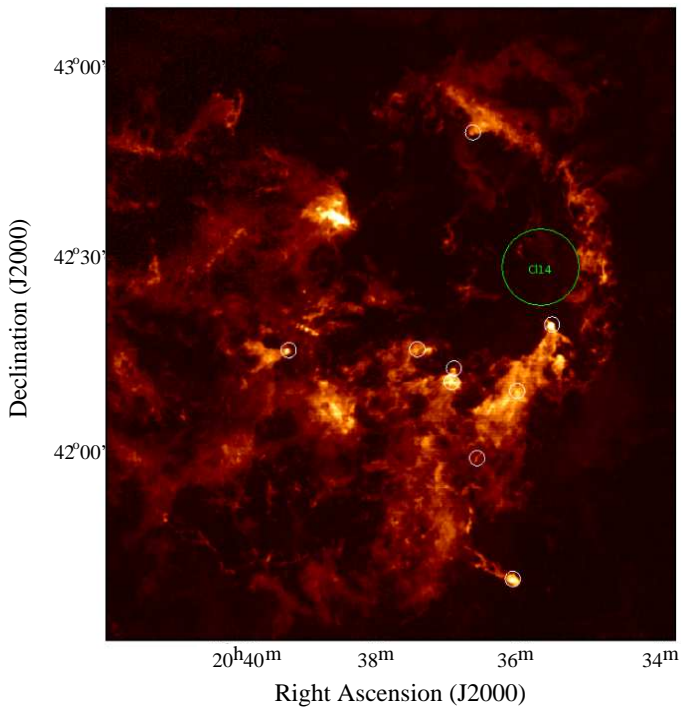}}
	\caption{Cometary features in a zeroth moment map of our Pathfinder. The \element[][12]{CO}{3--2} 
	emission was integrated from +10 to +25~km\,s$^{-1}$. Most cometary features point to the stellar
	cluster Cl\,14 -- which contains 12 OB-type stars -- indicating triggered star formation.}
	\label{fig:trig}
\end{figure}

A common occurrence around a stellar cluster containing massive stars
is thought to be triggered star formation.  Newly formed stars are
often found in evaporating gas globules or \emph{EGGs}
\citep{hest05}. These EGGs are located around the cluster's rim where
the UV radiation from the cluster has had time to sweep up the local
ISM and start to expose the young stellar objects within.  

The Pathfinder area is very close to the edge of the OB
association Cyg\,OB\,2 \citep{munc53}.  From 2\,MASS data
\citet{knod00} finds that the Cyg\,OB\,2 association more closely
resembles a globular cluster with a population of around $2600$
stars. He also finds that the diameter of Cyg\,OB\,2 is around
$4\degr$, almost double the original estimate. This allows for the
possibility that some embedded O- or B-type stars, not seen or not
included in the original Cyg\,OB\,2 association, are releasing large
amounts of energy into the Pathfinder area. 

Two comet-shaped nebulae pointing exactly towards the centre of
Cygnus~OB2 are present in our \element[][12]{CO}(3--2) Pathfinder
data. The large distances of these objects from the centre of Cyg~OB2,
one of them is over two degrees away, support the idea that the
Cyg\,OB\,2 association is larger than previously thought, although
this is not conclusive evidence.  Both cometary features contain
molecular outflows. One of the cometary objects is related to
IRAS~20294+4255 (see Figs.~\ref{fig:ofsample} and
\ref{fig:ofprof}). It contains the outflows G\,81.424+2.140 and
G\,81.435+2.147 (see Table~\ref{tab:of}) with a central velocity of
about $-2$~km\,s$^{-1}$.  This cometary cloud is located 
about 50~pc from the centre of Cygnus~OB2 projected to the
plane of the sky, assuming a distance of 1.7~kpc.  The
\element[][12]{CO}(3--2) emission from this feature does not show any
dynamical connection to DR~21 even though the radial velocity range is
very similar.  The other cometary feature pointing towards the centre
of Cygnus~OB2 contains outflow G\,80.832+0.570 at a velocity of
$+11.5$~km\,s$^{-1}$ which seems to be related to
IRAS~20343+4129. Whether there is a dynamical connection to W\,75\,N,
DR~17, and connected molecular clouds, which have similar radial
velocities, is not entirely clear.

It is interesting to note that the only cometary features in the Pathfinder that seem to be connected to Cygnus~OB2 show
such a high difference in radial velocity. Clearly observations covering the entire area
possibly influenced by Cygnus~OB2 are required to solve this little mystery. With a bigger sample
of cometary features connected to Cygnus~OB2 we could map the entire area around this association
in velocity space.

Another 
series of cometary nebulae in the Pathfinder are found pointing to new cluster 
candidate Cl\,14 and maybe Cl\,12 \citep{ledu02} (see Fig.~\ref{fig:trig}). These clusters overlap spatially with 
\ion{H}{ii} region DR\,17 \citep{down66}, and are thought to be related to the Cyg\,OB\,2 
association. These cometary objects are the 
\textit{pillars} from \citet{schn06}. No suitable UV radiation source had previously been 
identified that was able to cause their shape. Both of these clusters are proposed to be young 
clusters related to Cyg\,OB\,2 with Cl\,12 containing $\sim37$ OB stars and Cl\,14 $\sim12$. 
Either cluster candidate should be able to supply the required radiation to expose these 
protostars. Furthermore, Cl\,12 or Cl\,14 may have caused the star formation seen in these 
cometary nebulae. 

Several of these cometary features containing EGGs are pointing at Cl\,14 (Fig.~\ref{fig:trig}). 
These EGGs are likely the result of UV radiation 
from nearby stars dissociating and ionizing the outer layers of the molecular cloud. This warm 
gas is then blown away by stellar winds exposing the denser layers underneath. \citet{grit09} 
have studied the effect that O-type stars have on the surrounding ISM. They find that as the 
ionization front from massive O stars penetrates deeper into the molecular cloud, 
gravitational instability is induced in the tips of the pillars, which leads to the production 
of protostars. The presence of multiple protostars around groups of O- or B-type stars cannot 
be a chance coincidence and thus provides direct evidence for triggered star
formation. As the ionized gas expands through the molecular clouds it uncovers protostars. These 
protostars are still in the accreting phase, so as the UV radiation removes the molecular 
material around the protostar, a lower mass star will form. By studying the protostars formed 
in the EGGs around clusters and determining masses involved in the outflows we get an 
indication of whether or not these objects are being deprived of their molecular material 
\citep{hest05}. Some of these objects could finish their accretion phase by the time the 
ionization front reaches the protostar. This depends on the 
nature of the UV source and the distance to the protostars. To accurately study the impact of 
strong UV sources on protostars a large sample of objects around various sources of strong 
UV radiation is required.  

\subsection{Overall Structure of the Cygnus X Region}
		
\begin{figure*}
   \centerline{
	\includegraphics[bb=20 65 570 530,width=9cm,clip]{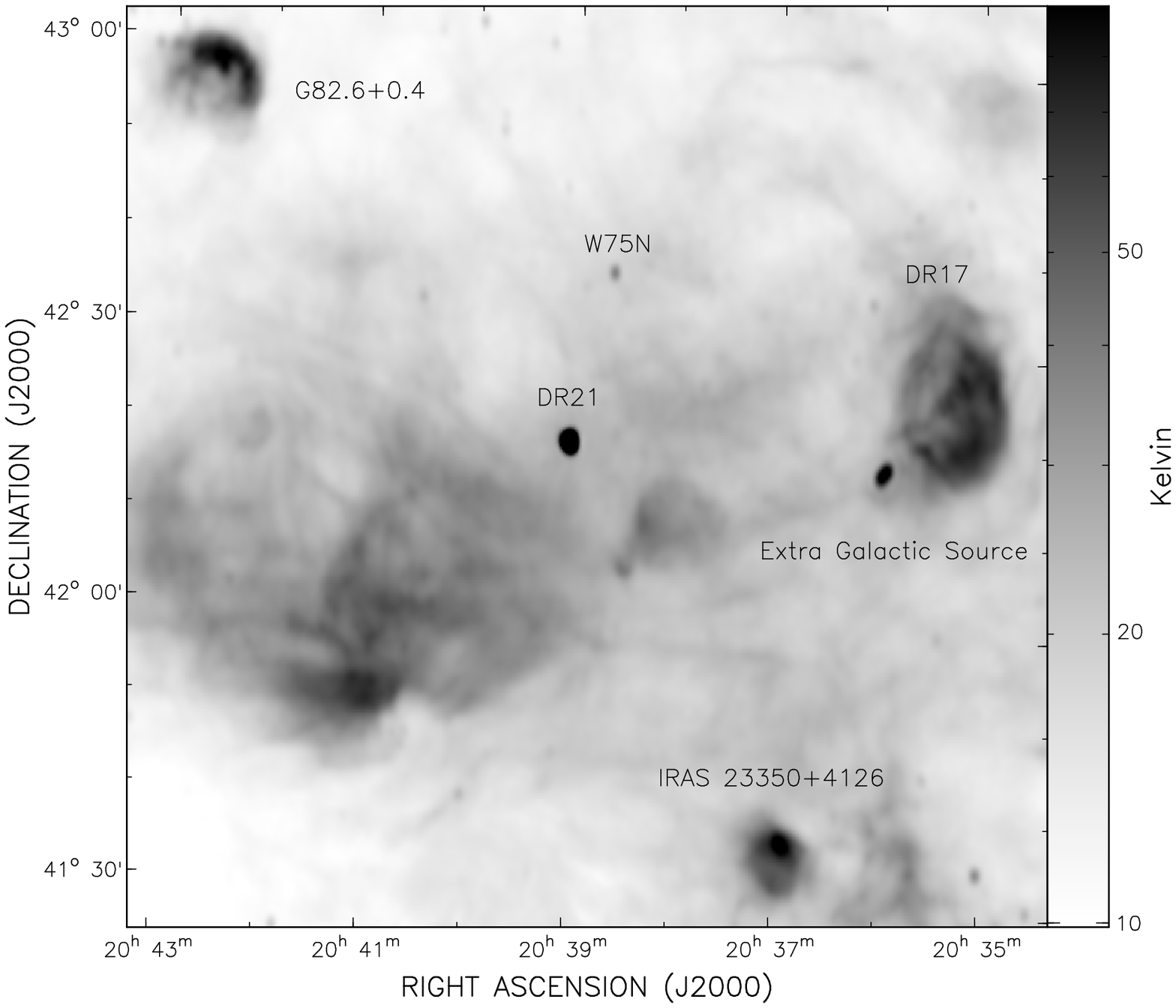}}
   \centerline{
   \begin{minipage}{8cm}
	\includegraphics[bb=50 50 530 500,width=8cm,clip]{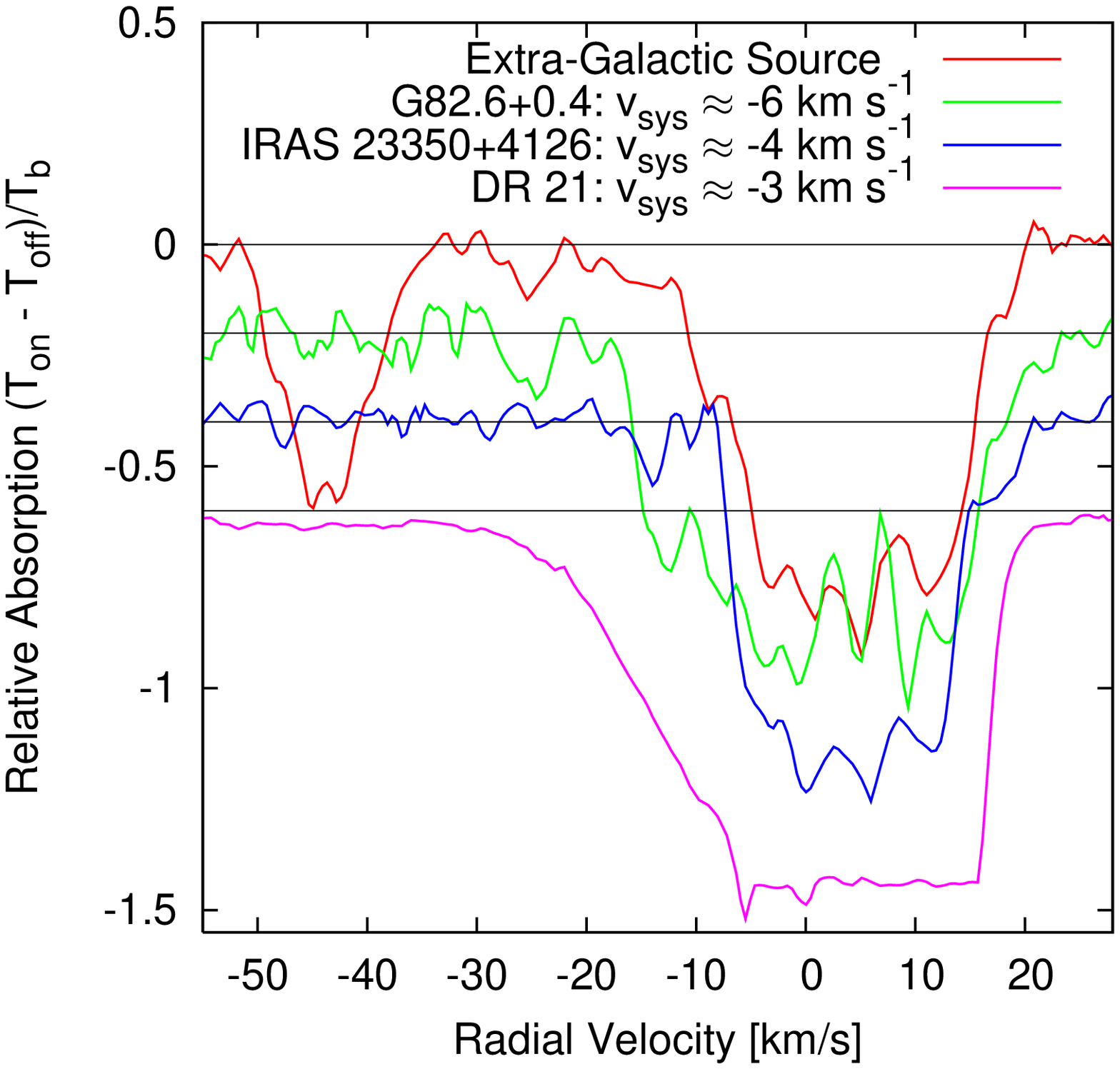}
   \end{minipage}
   \begin{minipage}{8cm}
	\includegraphics[bb=50 50 530 490,width=8cm,clip]{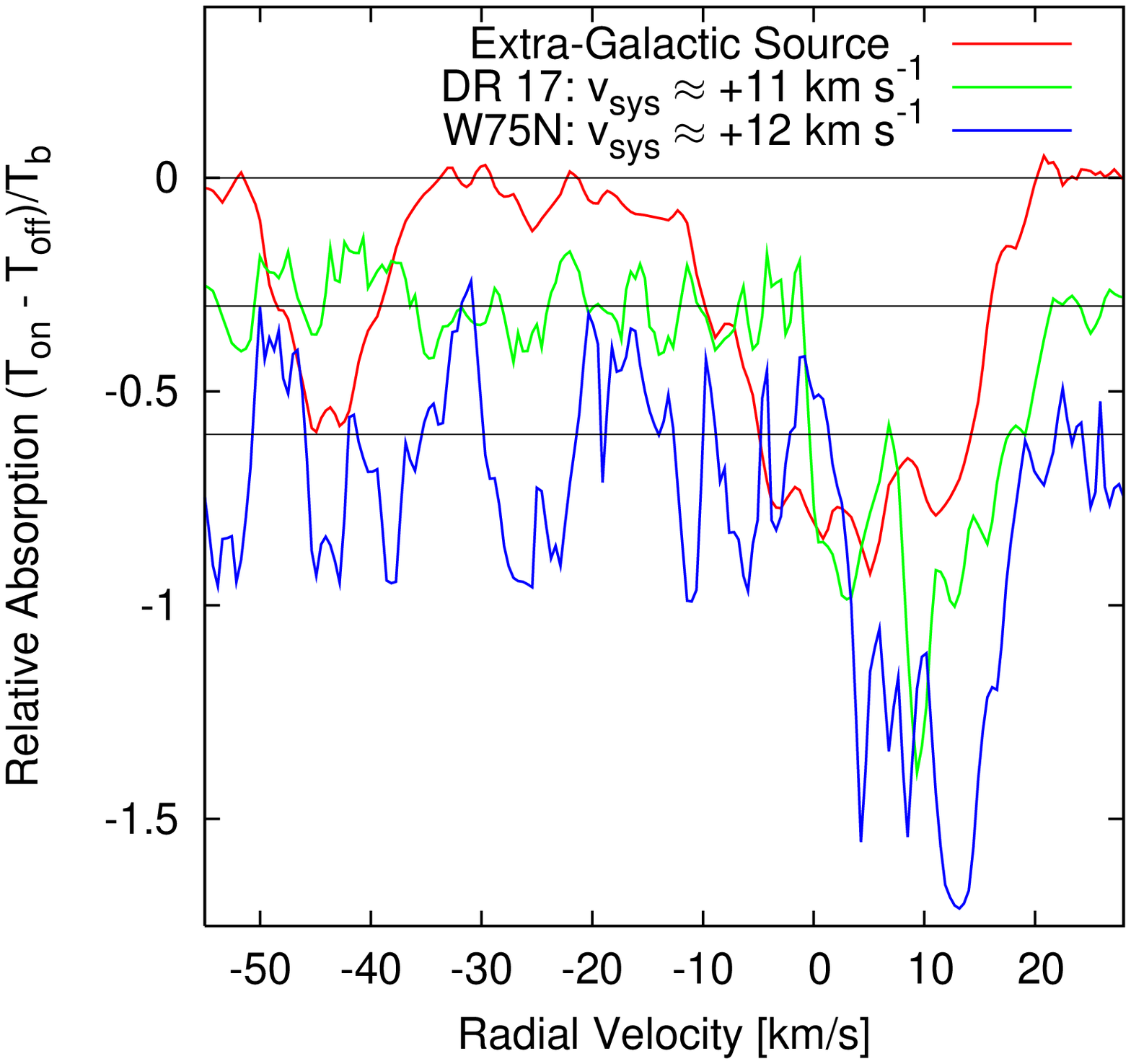}
   \end{minipage}}
	\caption{Radio continuum image taken from the Canadian
	Galactic Plane Survey at 1420~MHz (top) and \ion{H}{i}
	absorption profiles (bottom) of \ion{H}{ii} regions with high
	radio surface brightness in comparison with the profile of an
	extragalactic source. It is easy to see that the entire local
	\ion{H}{i} absorbs the radio continuum emission from those
	\ion{H}{ii} regions with a negative systemic velocity
	$v_{sys}$, while those with a high positive $v_{sys}$ are only
	absorbed by a part of that gas, indicating that the
	negative-velocity \ion{H}{ii} regions are farther away
	than the high positive-velocity ones. This rules out
	the possibility of them being part of one big dynamically
	connected structure.}
	\label{hiabs}
\end{figure*}

\citet{schn06} proposed that nearly all molecular clouds with radial
velocities between $-10$~km\,s$^{-1}$ and $+20$~km\,s$^{-1}$ in Cygnus X form 
groups that are dynamically connected, and that most of the Cygnus X objects 
are located at the distance of the Cyg OB2 cluster at 1.7~kpc. With the 
information gathered in the Pathfinder we will put the Schneider et al. postulate
to the test.

There are only two objects/features in the Cygnus~X region relevant for the Pathfinder which have well
determined distances and those are the Cygnus OB2 association at about 1.7~kpc
and the Great Cygnus Rift between 500 and 800~pc \citep{schn06}. The Great Cygnus Rift
is represented in the Pathfinder area by the low surface brightness molecular clouds
in the $^{12}$CO(3--2) emission layer at low positive velocities. The other two 
emission layers we found must be located behind the Cygnus Rift, simply because 
the Cygnus Rift is absorbing emission from all other molecular clouds. One example
is shown in Fig.~\ref{fig:radecvelo}: the red-shifted high velocity wings of DR~21
are absorbed by the faint layer related to the Cygnus Rift.
In addition we can see in Fig.~\ref{fig:radecvelo} (bottom) that the $^{12}$CO(3--2)
emission in the blue-shifted high velocity wings related to W\,75\,N are not absorbed by the 
dense gas at negative velocity related to DR~21. This implies that W\,75\,N is closer to 
us than DR~21. 

There is more evidence to support this. In Fig.~\ref{hiabs} we display \ion{H}{i} absorption
profiles of bright radio sources in the Pathfinder area and one nearby extragalactic source.
The \ion{H}{i} data were taken from the Canadian Galactic Plane Survey. The radio bright \ion{H}{ii}
regions DR~21, IRAS~20350+4126, and G82.6+0.4 all seem to be related to $^{12}$CO(3--2) emission
at negative radial velocities in our Pathfinder, while the \ion{H}{ii} regions DR~17 and W\,75\,N
are related to molecular gas at high positive velocity. In addition we computed a \ion{H}{i}
absorption profile towards a presumed extragalactic source next to DR~17 (indicated
in Fig.~\ref{hiabs}). The extragalactic source is the only source which shows an absorption
component at about $-45$~km\,s$^{-1}$, confirming that its location must be well behind the
other objects. 

DR~21, G82.6+0.4, and IRAS~20350+4126 have absorption profiles similar
to the extragalactic source at positive and low negative
velocities. DR~21 seem to have a high velocity wing towards negative
velocities, probably from dissociated and shocked molecular gas
accelerated towards us. This kind of signal has previously been
seen for other \ion{H}{ii} regions in the data of the Canadian
Galactic Plane Survey before \citep{koth02}. The sources with high
positive radial velocities, DR~17 and W\,75\,N show only absorption
components at high positive velocities, but nothing at low positive or
negative velocities. This is another proof that the sources related to
the $^{12}$CO(3--2) emission layer at high positive velocities must be
closer than those related to negative velocities.  We demonstrated in
Section 5.1 that both of these layers must have been pushed either
towards us or away from us, but since the more positive layer is
closer to us than the more negative one the source of the force that
pushed those layers could not have been the same. This casts doubt on
the proposed dynamical connection of most of the molecular clouds in
Cygnus~X.

\begin{figure}
   \begin{center}
	\includegraphics[bb=56 70 525 500,width=9cm]{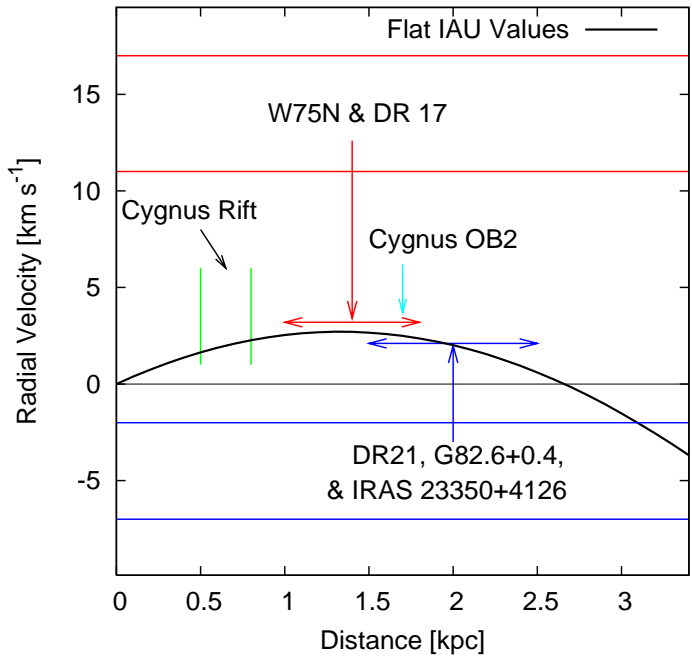}
   \end{center}
	\caption{Proposed locations of objects and emission layers along the line of sight
	as discussed in the text. The black line represents a flat rotation curve for the
	Galaxy with IAU supported values for $R_\odot$ and $v_\odot$.}
	\label{dist}
\end{figure}

Based on the information we have found in our data we propose the
following order along the line of sight towards Cygnus~X (as displayed
in Fig.~\ref{dist}). We have the two fixed distances for the Cygnus
Rift (500 to 800~pc) and Cygnus OB2 (1.7~kpc). Since the high
positive-velocity material must be pushed away from us we
believe its likely location is close to the tangent
point; there the energy required is lowest, because of the lower
velocity difference. This is also in agreement with a location between
the Cygnus Rift and the negative-velocity layer and a possible
connection to DR~17 \citep{ledu02}. Our line of sight within the local
spiral arm extends approximately to a distance of 2.5~kpc, as
can be seen in the Milky Way model by \citet{chur09}. We take this as
an upper limit. The sources at negative velocity also have to be
behind the highly positive layer and are therefore probably behind
Cygnus~OB2, although a connection cannot be ruled out.

\section{Conclusions}

The high critical density of the CO(3--2) transition makes it an
excellent tracer of dense molecular material. This has allowed us to
observe key stages in the molecule formation mechanism. We are
currently able to qualitatively discuss the properties of the cold
dense clouds making molecules, but a larger number of CO and HISA
correlations needs to be observed in order to quantify the molecule
formation rate. The Cygnus X region is also an excellent candidate for
this type of survey as it provides multiple generations of star
formation interacting at a relatively close distance from Earth. The
nature of the CO(3--2) line makes it an excellent tool for discovering
outflows. Of the outflows we detected more than 50\% are new
detections, and we have found all previously known outflows. In
order to probe the outflow engine more observations of optically thin
lines, for example \element[][13]{CO}(3--2) and C$^{18}$O(3--2), need
to be carried out. The detection of these outflows has provided direct
evidence of sequential star formation. By studying all of these
objects and the nature of their interaction with each other we can
develop a deep understanding of the properties of these complex
regions. Our Pathfinder has only observed around 15\% of the whole
Cygnus X region. With the success of this survey one can only wonder
what will come from a survey of the entire region.

\begin{acknowledgements}
The Dominion Radio Astrophysical Observatory is a National Facility operated by the National 
Research Council of Canada. The Canadian Galactic Plane Survey is a Canadian project with 
international partners, and is supported by the Natural Sciences and Engineering Research 
Council (NSERC). This research made use of data products from the Midcourse Space Experiment.  
Processing of the data was funded by the Ballistic Missile Defense Organization with 
additional support from NASA Office of Space Science.This research has made use of the NASA/ 
IPAC Infrared Science Archive, which is operated by the Jet Propulsion Laboratory, California 
Institute of Technology, under contract with the National Aeronautics and Space Administration.
\end{acknowledgements}

\bibliographystyle{aa}

\end{document}